\documentclass[prd,twocolumn,showpacs,superscriptaddress]{revtex4-1}
\usepackage{lipsum}
\usepackage{hyperref}
\usepackage{graphicx}
\usepackage{graphics}
\usepackage{amssymb}
\usepackage{amsthm}
\usepackage{mathbbol}
\usepackage{color}
\usepackage{amsfonts}
\usepackage{booktabs}
\usepackage{mathtools}
\usepackage{marginnote}
\usepackage[normalem]{ulem}

\newcommand{\be}{\begin{equation}}
\newcommand{\ee}{\end{equation}}
\newcommand{\bea}{\begin{eqnarray}}
\newcommand{\eea}{\end{eqnarray}}
 
\newcommand{\gtapprox}{\raisebox{-0.5ex}{$\,\stackrel{>}{\scriptstyle\sim}\,$}}
\newcommand{\ltapprox}{\raisebox{-0.5ex}{$\,\stackrel{<}{\scriptstyle\sim}\,$}}
\begin{document}

\title{Tetraquark interpolating fields in a lattice QCD investigation of the $D_{s0}^\ast(2317)$ meson}

\author{Constantia Alexandrou}
\affiliation{Department of Physics, University of Cyprus, P.O. Box 20537, 1678 Nicosia, Cyprus}
\affiliation{Computation-based Science and Technology Research Center, \\ The Cyprus Institute, 20 Kavafi Street, 2121 Nicosia, Cyprus}

\author{Joshua Berlin}
\affiliation{Goethe-Universit\"at Frankfurt am Main, Institut f\"ur Theoretische Physik, Max-von-Laue-Stra{\ss}e 1, D-60438 Frankfurt am Main, Germany}

\author{Jacob Finkenrath}
\affiliation{Computation-based Science and Technology Research Center, The Cyprus Institute, 20 Kavafi Street, 2121 Nicosia, Cyprus}

\author{Theodoros Leontiou}
\affiliation{Department of Mechanical Engineering, Frederick University, 1036 Nicosia, Cyprus}

\author{Marc Wagner}
\affiliation{Goethe-Universit\"at Frankfurt am Main, Institut f\"ur Theoretische Physik, Max-von-Laue-Stra{\ss}e 1, D-60438 Frankfurt am Main, Germany}

\begin{abstract}
We investigate the $D_{s0}^\ast(2317)$ meson using lattice QCD and considering correlation functions of several $\bar{c} s$ two-quark and $\bar{c} s (\bar{u} u + \bar{d} d)$ four-quark interpolating fields. These interpolating fields generate different structures in color, spin and position space including quark-antiquark pairs, tetraquarks and two-meson scattering states. For our computation we use an ensemble simulated with pion mass $m_\pi \approx 0.296 \, \textrm{GeV}$ and spatial volume of extent $2.90 \, \textrm{fm}$. We find in addition to the expected spectrum of two-meson scattering states another state around $60 \, \textrm{MeV}$ below the $D K$ threshold, which we interpret as the $D_{s0}^\ast(2317)$ meson. This state couples predominantly to a quark-antiquark interpolating field and only weakly to a $D K$ two-meson interpolating field. The coupling to the tetraquark interpolating fields is essentially zero, rendering a tetraquark interpretation of the $D_{s0}^\ast(2317)$ meson rather unlikely. Moreover, we perform a scattering analysis using L\"uscher's method and the effective range approximation to determine the $D_{s0}^\ast(2317)$ mass for infinite spatial volume. We find this mass $51 \, \textrm{MeV}$ below the $D K$ threshold, rather close to both our finite volume result and the experimentally observed value.
\end{abstract}

\maketitle



\section{\label{SEC008}Introduction}

The $D_{s0}^\ast(2317)$ meson with quantum numbers $I(J^P) = 0(0^+)$, strangeness $S = \pm 1$ and charm $C = S$ has mass $m_{D_{s0}^\ast} = 2.3178(5) \, \textrm{GeV}$, around $45 \, \textrm{MeV}$ below the $D K$ threshold \cite{Aubert:2003fg,Besson:2003cp,Krokovny:2003zq,Tanabashi:2018oca}. This experimental result is in contrast to theoretical predictions from quark models (see e.g.\ Refs.\ \cite{Godfrey:1985xj,Godfrey:1986wj,Ebert:2009ua}), where the $D_{s0}^\ast(2317)$ meson is treated as a $\bar{c} s$ quark-antiquark pair, which leads to a significantly larger mass in the range of $100 \, \textrm{MeV}$ to $200 \, \textrm{MeV}$ above the experimental value. Because of that discrepancy, there is an ongoing debate about the quark composition of the $D_{s0}^\ast(2317)$ meson. Besides a standard quark-antiquark structure it could also have a four-quark structure. For example Refs.\ \cite{Maiani:2004vq,Bracco:2005kt,Dmitrasinovic:2005gc} propose a tetraquark structure, while Ref.\ \cite{Ebert:2010af} provides arguments against such a scenario. Another possibility is a $D K$ mesonic molecule structure as e.g.\ suggested by Refs.\ \cite{Barnes:2003dj,Chen:2004dy}. This picture is also supported by recent papers \cite{Du:2017zvv,MartinezTorres:2017bdo,Albaladejo:2018mhb,Guo:2018tjx}, where $D K$ molecular components from around $60 \%$ to $75 \%$ are found. Other interesting approaches, which are able to explain the surprisingly low mass of the $D_{s0}^\ast(2317)$ meson, are e.g.\ presented in Ref.\ \cite{vanBeveren:2003kd}, where the $D_{s0}^\ast(2317)$ meson is a standard $\bar{c} s$ configuration with the coupling to the nearby $D K$ threshold taken into account, and in Refs.\ \cite{Kolomeitsev:2003ac,Hofmann:2003je,Guo:2018kno}, which are based on an SU(3) chiral Lagrangian. For a more detailed discussion of the properties of the $D_{s0}^\ast(2317)$ meson and existing literature we refer to the review articles \cite{Zhu:2007wz,Chen:2016spr}.

Early quenched lattice QCD studies \cite{Boyle:1997aq,Boyle:1997rk,Lewis:2000sv,Hein:2000qu,Bali:2003jv,Dougall:2003hv,diPierro:2003iw} of the $D_{s0}^\ast(2317)$ meson found masses significantly larger than the experimental result, similar to quark model predictions. There are also more recent lattice QCD studies \cite{Mohler:2011ke,Namekawa:2011wt,Bali:2011dc,Bali:2012ua,Moir:2013ub,Kalinowski:2015bwa,Cichy:2016bci,Cheung:2016bym,Chen:2017kxr}, where only quark-antiquark interpolating fields of flavor structure $\bar{c} s$ were taken into account. The majority of these studies also finds masses for the $D_{s0}^\ast(2317)$ meson, which are larger than the experimental value, in particular if extrapolations to physical quark masses and to the continuum were performed (see e.g.\ Ref.\ \cite{Cichy:2016bci}). If, however, in addition to quark-antiquark interpolating fields also two-meson $D K$ interpolating fields are included, as done in the recent precision lattice QCD computations presented in Refs.\ \cite{Mohler:2013rwa,Lang:2014yfa,Bali:2017pdv}, $D_{s0}^\ast(2317)$ masses below the $D K$ threshold and close or consistent with the experimental result are found. In these investigations almost physical $u$ and $d$ quark masses were used, corresponding to pion masses $m_\pi \approx 0.156 \, \textrm{GeV}$ and $m_\pi \approx 0.150 \, \textrm{GeV}$, respectively, and L\"uscher's method was employed, to obtain the meson mass at infinite spatial volume. In this context it is also interesting to mention two closely related lattice QCD investigations. In Ref.\ \cite{Liu:2012zya} scattering of charmed and light pseudoscalar mesons was studied, including $D \bar{K}$ scattering, and by using SU(3) flavor symmetry the $D_{s0}^\ast(2317)$ mass was obtained in agreement with experiment and support for the interpretation as a $D K$ molecule was found. In Ref.\ \cite{Moir:2016srx} scattering in the $D_0^\ast$ sector was studied, however, with rather heavy quark masses, somewhere between the physical light and strange quark, corresponding to a pion mass $m_\pi \approx 0.391 \ \textrm{GeV}$, which led to some insights on the qualitative difference between the $D_0^\ast(2400)$ meson and the $D_{s0}^\ast(2317)$ meson.

In this work we also study the $D_{s0}^\ast(2317)$ meson using lattice QCD with particular focus on tetraquark interpolating fields. As in the aforementioned lattice QCD investigations \cite{Mohler:2013rwa,Lang:2014yfa,Bali:2017pdv} we consider both quark-antiquark and two-meson interpolating fields. In addition, we include for the first time also tetraquark interpolating fields, where the four quarks are centered at the same point in space. We implemented color and spin contractions, where two standard meson interpolating fields of quark-antiquark type are put on top of each other (resembling $D K$ and $D_s \eta$ mesonic molecules), as well as contractions, which have a diquark-antidiquark structure. Including such tetraquark interpolating fields might be essential, as it has recently been reported in Ref.\ \cite{Darvish:2019oie} for the positive parity mesons $a_0(980)$ and $K_0^\ast(700)$. In both cases a low-lying energy level is missed, if they are not taken into account. Moreover, we compute the couplings of the low lying states to different types of two-quark and four-quark interpolating fields and compare the spectra obtained from different subsets of interpolating fields. This might shed additional light on the question, whether the $D_{s0}^\ast(2317)$ meson is predominantly a $\bar{c} s$ state or rather has a large tetraquark component.

We perform our computations in a single spatial volume of extent $2.90 \, \textrm{fm}$ and at unphysically heavy $u$ and $d$ quark mass corresponding to $m_\pi \approx 296 \, \textrm{MeV}$. For the analysis of correlation functions we apply the Athens Model Independent Analysis Scheme (AMIAS), an analysis method based on statistical concepts for extracting excited states from correlation functions. AMIAS is a novel analysis method, which has previously been used in a study of the nucleon spectrum and the $a_0(980)$ meson \cite{Alexandrou:2014mka,Alexandrou:2017itd}. AMIAS utilizes all the information encoded in the correlation function with the particular advantage of exploiting also data at small temporal separations, where statistical errors are typically small. In addition to AMIAS we also use the standard generalized eigenvalue problem (GEVP) method, i.e.\ we solve generalized eigenvalue problems and extract the spectrum from effective energy plateaus (cf.\ e.g.\ \cite{Blossier:2009kd} and references therein). Note that both the GEVP and AMIAS provide information on the relative importance of the considered interpolating fields. Combining both methods allows to check the robustness of our results.

This paper is organized as follows: In section~\ref{sec:simu} we describe the lattice setup and techniques with particular focus on the implemented interpolating fields. In section~\ref{sec:euccor} we discuss the spectral decomposition of the corresponding two-point correlation functions. A short description of our two analysis methods, the GEVP and AMIAS, is provided in section~\ref{SEC467}. Section~\ref{sec:cormat} is the main section of this work, where our numerical results are presented. First, in section~\ref{SEC455}, we show several finite volume spectra for the sector with $D_{s0}^\ast(2317)$ quantum numbers corresponding to different sets of interpolating fields. Based on these results the importance of each interpolating field is discussed. Then, in section~\ref{SEC762}, we perform a scattering analysis using L\"uscher's method and the effective range expansion to determine the $D_{s0}^\ast(2317)$ mass at infinite volume. In section~\ref{sec:conclusions} we summarize our findings and give our conclusions.


\section{\label{sec:simu}Interpolating fields and lattice setup}

To investigate the $D_{s0}^\ast(2317)$ meson, we consider a $7 \times 7$ correlation matrix
%
%
%
\begin{eqnarray}
\label{EQN789} C_{j k}(t) = \Big\langle \mathcal{O}^j(t_2) \mathcal{O}^{k \dagger}(t_1) \Big\rangle \quad , \quad t = t_2-t_1 .
\end{eqnarray}
The interpolating fields $\mathcal{O}^j$, $j=1,\ldots,7$ have either a two-quark $\bar{c} s$ or a four-quark $\bar{c} s \bar{q} q$ structure, where $\bar{q} q = (\bar{u} u + \bar{d} d) / \sqrt{2}$. In detail we consider the interpolating fields
\begin{eqnarray}
\label{EQN002} & & \mathcal{O}^1 = \mathcal{O}^{q \bar{q}, \ 1} = N_1 \sum_{\bf{x}} \Big({\bar c}({\bf x}) s({\bf x})\Big) \\
\label{EQN002_} & & \mathcal{O}^2 = \mathcal{O}^{q \bar{q}, \ \gamma_0} = N_2 \sum_{\bf{x}} \Big({\bar c}({\bf x}) \gamma_0 s({\bf x})\Big) \\
\nonumber & & \mathcal{O}^3 = \mathcal{O}^{D K, \ \textrm{point}} \\
\label{EQN007} & & \hspace{0.7cm} = N_3 \sum_{\bf{x}} \Big({\bar c}({\bf x}) \gamma_5 q({\bf x})\Big) \Big({\bar q}({\bf x}) \gamma_5 s({\bf x})\Big) \\
\nonumber & & \mathcal{O}^4 = \mathcal{O}^{D_s \eta, \ \textrm{point}} \\
\label{EQN117} & & \hspace{0.7cm} = N_4 \sum_{\bf{x}} \Big({\bar c}({\bf x}) \gamma_5 s({\bf x})\Big) \Big({\bar q}({\bf x}) \gamma_5 q({\bf x})\Big) \\
\nonumber & & \mathcal{O}^5 = \mathcal{O}^{Q \bar{Q}, \, \gamma_5} \\
\nonumber & & \hspace{0.7cm} = N_5 \sum_{\bf{x}} \epsilon_{a b c} \Big({\bar c}_b({\bf x}) (C \gamma_5) {\bar q}_c^T({\bf x})\Big) \\
\label{EQN631} & & \hspace{1.4cm} \epsilon_{a d e} \Big(q_d^T({\bf x}) (C \gamma_5) s_e({\bf x})\Big) \\
\nonumber & & \mathcal{O}^6 = \mathcal{O}^{D K, \ \textrm{2part}} \\
\label{EQN113} & & \hspace{0.7cm} = N_6 \sum_{{\bf x},{\bf y}} \Big({\bar c}({\bf x}) \gamma_5 q({\bf x})\Big) \Big({\bar q}({\bf y}) \gamma_5 s({\bf y})\Big) \\
\nonumber & & \mathcal{O}^7 = \mathcal{O}^{D_s \eta, \ \textrm{2part}} \\
\label{EQN003} & & \hspace{0.7cm} = N_7 \sum_{{\bf x},{\bf y}} \Big({\bar c}({\bf x}) \gamma_5 s({\bf x})\Big) \Big({\bar q}({\bf y}) \gamma_5 q({\bf y})\Big) .
\end{eqnarray}
$C$ denotes the charge conjugation matrix and the normalization factors $N_j$ are chosen such that $C_{j j}(t=a) = 1$ (no sum over $j$; $a$ is the lattice spacing), i.e.\ in a way that the interpolating fields generate trial states with similar norm. All interpolating fields couple to the $D_{s0}^\ast(2317)$ meson and to other states with the same quantum numbers. As in previous lattice QCD computations \cite{Mohler:2013rwa,Lang:2014yfa,Bali:2017pdv} we consider quark-antiquark interpolating fields, $\mathcal{O}^{q \bar{q}, \ 1}$ and $\mathcal{O}^{q \bar{q}, \ \gamma_0}$, as well as two-meson interpolating fields, $\mathcal{O}^{D K, \ \textrm{2part}}$ and $\mathcal{O}^{D_s \eta, \ \textrm{2part}}$. In Refs.\ \cite{Mohler:2013rwa,Lang:2014yfa,Bali:2017pdv} it was shown that the latter interpolating fields are essential to determine the energy of the ground state and the first excitation reliably. In addition we implemented the tetraquark interpolating fields $\mathcal{O}^{D K, \ \textrm{point}}$, $\mathcal{O}^{D_s \eta, \ \textrm{point}}$ and $\mathcal{O}^{Q \bar{Q}, \, \gamma_5}$ with the four quark operators located at the same point in space.

The interpolating fields $\mathcal{O}^{D K, \ \textrm{2part}}$ and $\mathcal{O}^{D_s \eta, \ \textrm{2part}}$ mostly generate $D K$ and $D_s \eta$ scattering states, which are expected to have energies somewhat above the mass of the $D_{s0}^\ast(2317)$ meson ($m_D + m_K - m_{D_{s0}^\ast} \approx 45 \, \textrm{MeV}$ and $m_{D_s} + m_\eta - m_{D_{s0}^\ast} \approx 200 \, \textrm{MeV}$ \cite{Tanabashi:2018oca}). In contrast to $\mathcal{O}^{D K, \ \textrm{2part}}$ and $\mathcal{O}^{D_s \eta, \ \textrm{2part}}$, where both mesons have zero momentum, the interpolating fields $\mathcal{O}^{D K, \ \textrm{point}}$ and $\mathcal{O}^{D_s \eta, \ \textrm{point}}$ represent two mesons centered at the same point in space and, thus, resemble mesonic molecules. Similarly, due to the different color structure, $\mathcal{O}^{Q \bar{Q}, \, \gamma_5}$ resembles a diquark-antidiquark pair.

Tetraquark interpolating fields like $\mathcal{O}^{D K, \ \textrm{point}}$, $\mathcal{O}^{D_s \eta, \ \textrm{point}}$ and $\mathcal{O}^{Q \bar{Q}, \, \gamma_5}$ were not considered in previous lattice QCD studies of the $D_{s0}^\ast(2317)$ meson. Thus, the main goal of this work is to explore, whether the inclusion of these tetraquark interpolating fields has an effect on the lattice QCD determination of the low-lying spectrum. Similar recent investigations of systems not including the $D_{s0}^\ast(2317)$ meson have led to different findings regarding the importance of tetraquark interpolating fields. While in Refs.\ \cite{Cheung:2017tnt,Alexandrou:2017itd} only marginal differences in the resulting spectra of the $I=1$ hidden-charm and doubly-charmed sectors and the $a_0(980)$ sector were found, Ref.\ \cite{Darvish:2019oie} observed additional energy levels both with $K_0^\ast(700)$ and $a_0(980)$ quantum numbers. In this work, we also compute and compare the overlaps of the corresponding trial states $\mathcal{O}^j | \Omega \rangle$ ($| \Omega \rangle$ denotes the vacuum) to the lowest energy eigenstate, to obtain certain information about the quark composition of the $D_{s0}^\ast(2317)$ meson. This might contribute to the ongoing debate, whether the $D_{s0}^\ast(2317)$ meson is predominantly a quark-antiquark pair or a tetraquark (see the discussion in section~\ref{SEC008}).

Note that the interpolating fields $\mathcal{O}^3$ to $\mathcal{O}^7$ do not generate orthogonal trial states. For example the terms with $\mathbf{x} = \mathbf{y}$ in Eqs.\ (\ref{EQN113}) and (\ref{EQN003}) also appear in Eqs.\ (\ref{EQN007}) and (\ref{EQN117}). Similarly, one can relate two-meson combinations to diquark-antidiquark combinations via a Fierz identity, i.e.\ some of the terms present in Eqs.\ (\ref{EQN007}) and (\ref{EQN117}) are also part of Eq.\ (\ref{EQN631}) and vice versa. Even though the seven interpolating fields do not generate orthogonal trial states, they are not linearly dependent either, because each of them contains terms not present in any of the other six. Their non-orthogonality does not cause any particular problems during our analyses, because the two methods we use, the GEVP and AMIAS, are both able to deal with correlation matrices based on non-orthogonal trial states. We remark that on a technical level this work is similar to our lattice QCD investigation of the $a_0(980)$ meson \cite{Alexandrou:2017itd}, because to a large extent the same interpolating fields are used, just with different quark flavors.

To increase the coupling of the interpolating fields to the low-lying energy eigenstates, quark fields in Eqs.\ (\ref{EQN002}) to (\ref{EQN003}) are Gaussian smeared with APE smeared spatial gauge links (cf.\ Refs.\ \cite{Albanese:1987ds,Gusken:1989qx}). The smearing parameters are $\kappa_\textrm{Gauss} = 0.5$, $N_\textrm{Gauss} = 50$, $\alpha_\textrm{APE} = 0.45$ and $N_\textrm{APE} = 20$, where detailed equations are given in \cite{Jansen:2008si}.

To compute the correlation functions, we use an ensemble of around 500 gauge link configurations generated with $N_f=2+1$ dynamical Wilson clover quarks and the Iwasaki gauge action by the PACS-CS Collaboration \cite{Aoki:2008sm}. The lattice size is $64 \times 32^3$ with lattice spacing $a = 0.0907(14) \, \textrm{fm}$, i.e.\ the spatial lattice extent $L$ is around $2.90 \, \textrm{fm}$. The $u$ and $d$ quark mass and the $s$ quark mass correspond to the pion mass $m_\pi \approx 0.296 \, \textrm{GeV}$ and the kaon mass $m_K \approx 0.597 \, \textrm{GeV}$, i.e.\ are both heavier than in the real world, while the $c$ quark mass corresponds to the $D$ meson mass $m_D \approx 1.845 \, \textrm{GeV}$, i.e.\ is slightly lighter (see the detailed discussion in section~\ref{SEC762}). Note that the $c$ quark only appears as a valence quark.

In a recent publication \cite{Abdel-Rehim:2017dok}, we implemented and compared various combinations of techniques for the computation of propagators and correlation functions including point-to-all propagators, stochastic timeslice-to-all propagators, the one-end trick and sequential propagators. For each diagram of a similar $6 \times 6$ correlation matrix, which we used to study the $a_0(980)$ meson \cite{Alexandrou:2017itd}, we determined the most efficient combination of techniques. We have applied the same combinations of techniques in this work to compute the $7 \times 7$ correlation matrix (\ref{EQN789}) with the interpolating fields (\ref{EQN002}) to (\ref{EQN003}). Finding efficient methods is particularly important for diagrams, where quarks propagate within a timeslice, e.g.\ diagrams containing closed quark loops. These diagrams are significantly more noisy than their counterparts, where quarks do not propagate within a timeslice. Their noise-to-signal ratio grows exponentially with increasing temporal separation as discussed in Ref.\ \cite{Abdel-Rehim:2017dok}.


\section{\label{sec:euccor}Correlation functions for periodic temporal direction}

A correlation function computed on a lattice with periodic temporal direction of extension $T$ can be expanded according to
\begin{eqnarray}
\nonumber & & C_{j k}(t) = \Big\langle \mathcal{O}^j(t) \mathcal{O}^{k \dagger}(0) \Big\rangle \\
\label{EQN566} & & \hspace{0.7cm} = \frac{1}{Z} \sum_{m,n} e^{-E_m (T-t)} c_{m,n}^j e^{-E_n t} (c_{m,n}^k)^\ast
\end{eqnarray}
with energy eigenstates $| m \rangle$, corresponding energy eigenvalues $E_m$, possibly complex $c_{m,n}^j = \langle m | \mathcal{O}^j | n \rangle$ and $Z = \sum_m e^{-E_m T}$.

Using the QCD symmetries charge conjugation and time reversal one can show that all elements of the correlation matrix (\ref{EQN789}) with interpolating fields (\ref{EQN002}) to (\ref{EQN003}) are real. Moreover, one can rewrite Eq.\ (\ref{EQN566}) in more convenient form,
\begin{eqnarray}
\nonumber & & C_{j k}(t) = \frac{1}{Z} \sum_{m,n} e^{-(E_m + E_n) T/2} c^j_{m,n} c^k_{m,n} \\
\label{eq:cosh} & & \hspace{1.4cm} H_{j k}((E_m - E_n) (t - T/2)) ,
\end{eqnarray}
with real $c^j_{m,n}$ and
\begin{eqnarray}
\nonumber & & H_{j k}(x) \\
\nonumber & & \hspace{0.7cm} = \left\{\begin{array}{cc}
-\sinh(x) & \textrm{for } j = 2 , k \neq 2 \textrm{ and } j \neq 2 , k = 2 \\
+\cosh(x) & \textrm{otherwise}
\end{array}\right. . \\
\label{eq:cosh_} & &
\end{eqnarray}
Since the elements of the correlation matrix are either symmetric with respect to the reversal of time, $C_{j k}(t) = +C_{j k}(T - t)$ for $H_{j k}(x) = +\cosh(x)$, or antisymmetric, $C_{j k}(t) = -C_{j k}(T - t)$ for $H_{j k}(x) = -\sinh(x)$, it is sufficient to restrict the following discussion to temporal separations $0 \leq t \leq T/2$.

For sufficiently large $T$, where $Z \approx e^{-E_\Omega T}$ ($\Omega$ denotes the vacuum), and for sufficiently large $t$, Eq.\ (\ref{eq:cosh}) reduces to
\begin{eqnarray}
\nonumber & & C_{j k}(t) = \sum_m^\textrm{truncated} 4 e^{-\mathcal{E}_m T/2} c^j_{m,\Omega} c^k_{m,\Omega} H_{j k}(\mathcal{E}_m (t - T/2)) , \\
\label{eq:cosh_fit} & &
\end{eqnarray}
if the correlation function is not contaminated by effects related to multi-hadron states as discussed below. $\mathcal{E}_m = E_m - E_\Omega$ and $\sum_m^\textrm{truncated}$ denotes the sum over a finite number of low-lying energy eigenstates in the sector with $D_{s0}^\ast(2317)$ quantum numbers, which is probed by the interpolating fields (\ref{EQN002}) to (\ref{EQN003}) (in the following we assume the ordering $\mathcal{E}_0 \leq \mathcal{E}_1 \leq \mathcal{E}_2 \leq \ldots$).

For temporal separations $t$ around $T/2$ the correlation functions $C_{j k}(t)$ have a more complicated expansion than Eq.\ (\ref{eq:cosh_fit}), if there are low-lying multi-hadron states with the same quantum numbers. In our case, i.e.\ for $D_{s0}^\ast(2317)$ quantum numbers, the lowest multi-hadron state is a $D K$ scattering state, which has an energy only slightly above the mass of the $D_{s0}^\ast(2317)$ meson. Clearly, the interpolating fields (\ref{EQN002}) to (\ref{EQN003}) do not only excite such a $D K$ scattering state, when applied to the vacuum $| \Omega \rangle$, but also yield non-vanishing matrix elements $\langle D | \mathcal{O}^j | K \rangle$ and $\langle K | \mathcal{O}^j | D \rangle$, i.e.\ annihilate a kaon and create a $D$ meson and vice versa. For example a significant contribution to $C_{j j}(t)$ is
\begin{eqnarray}
\nonumber & & \frac{2}{Z} e^{-(E_D + E_K) T/2} (c^j_{D,K})^2 \textrm{cosh}((E_D - E_K) (t - T/2)) \\
\nonumber & & \hspace{0.7cm} \approx 2 e^{-(m_D + m_K) T/2} (c^j_{D,K})^2 \\
\label{EQN620} & & \hspace{1.4cm} \textrm{cosh}((m_D - m_K) (t - T/2))
\end{eqnarray}
as can be seen from Eq.\ (\ref{eq:cosh}). Assuming coefficients $|c^j_{D,K}| \approx |c^j_{D K,\Omega}|$, where $D K$ denotes a low-lying $D K$ scattering state, one can see that in the region of $t \approx T/2$ the corresponding terms in Eq.\ (\ref{eq:cosh}) are comparable in magnitude. Therefore, terms as in Eq.\ (\ref{EQN620}) have to be taken into account, when extracting energy levels from correlation functions at large temporal separations $t \approx T/2$. For smaller temporal separations $t$ such contributions may be neglected, since they are exponentially suppressed $\propto e^{-2 m_K (t - T/2)}$ with decreasing $t$. Analytical estimates as well as numerical experiments have shown, that within our setup this is the case for $t \ltapprox 15 \, a$, which is an upper bound for all $t$ fitting ranges used in the following.


\section{\label{SEC467}Analysis methods}

To analyze the $7 \times 7$ correlation matrix discussed in section~\ref{sec:euccor} and various submatrices, we use both the GEVP method and the AMIAS method. While the GEVP method is quite common and very well known, the AMIAS method has proven to be particularly suited to study excited states \cite{Alexandrou:2014mka} and was succesfully used in our related previous lattice QCD study of the $a_0(980)$ meson \cite{Alexandrou:2017itd}. In section~\ref{sec:cormat} we will show that both methods yield consistent results, which we consider to be an important cross-check, in particular due to the fact that the signal-to-noise ratios of the elements of the correlation matrix grow rapidly with increasing temporal separations. In the following we summarize both methods and discuss the details of our analyses.


\subsection{\label{SEC496}GEVP method}

A commonly used method to extract several energy levels from an $N \times N$ correlation matrix is to solve the generalized eigenvalue problem
\begin{eqnarray}
\label{gevp2} C(t) \mathbf{v}_m(t,t_0) = \lambda_m(t,t_0) C(t_0) \mathbf{v}_m(t,t_0)
\end{eqnarray}
(see e.g.\ Ref.\ \cite{Blossier:2009kd} and references therein), where $C(t)$ is the correlation matrix with entries $C_{j k}(t)$ ($j,k=1,\ldots,N$), $\mathbf{v}_m(t,t_0)$ the eigenvector corresponding to the eigenvalue $\lambda_m(t,t_0)$ ($m=0,\ldots,N-1$) and $t_0 \geq a$ an input parameter. We use $t_0 = a$, which is a typical choice.

A number of $N$ effective energies $E_{\textrm{eff},m}(t)$ can be obtained by solving
\begin{eqnarray}
\label{eq:lambdaeffEn} \frac{\lambda_m(t,t_0)}{\lambda_m(t-a,t_0)} = \frac{\textrm{cosh}(E_{\textrm{eff},m}(t) (t - T/2))}{\textrm{cosh}(E_{\textrm{eff},m}(t) (t-a - T/2))}
\end{eqnarray}
for each eigenvalue $\lambda_m(t,t_0)$. At sufficiently large, but not too large temporal separations, i.e.\ in a $t$-region, where Eq.\ (\ref{eq:cosh_fit}) is a valid parameterization of the correlation matrix, the effective energies $E_{\textrm{eff},m}(t)$ exhibit plateaus. The values of these plateaus correspond to the $N$ lowest energy levels in the sector probed by the interpolating fields, i.e.\ to $\mathcal{E}_m$. We determine each energy level $\mathcal{E}_m$ by first fitting
\begin{eqnarray}
\nonumber & & f(t) = A_0 \cosh({E}_0 (t - T/2)) + A_1 \cosh(E_1 (t - T/2)) \\
 & &
\end{eqnarray}
to the eigenvalue $\lambda_m(t,t_0)$ in the region $t_\textrm{min} \leq t \leq t_\textrm{max}$, where $A_0$, $A_1$ and $E_0 < E_1$ are fitting parameters. $t_\textrm{min}$ and $t_\textrm{max}$ are chosen as follows:
\begin{itemize}
\item $t_\textrm{min}$ is the smallest temporal separation $t$, where
\begin{eqnarray}
 & & \Big|E_{\textrm{eff},m}^f(t) - E_0\Big| \leq \Delta E_{\textrm{eff},m}(t)
\end{eqnarray}
($E_{\textrm{eff},m}^{f}(t)$ is the solution of
\begin{eqnarray}
\frac{f(t)}{f(t-a)} = \frac{\textrm{cosh}(E_{\textrm{eff},m}^f(t) (t - T/2))}{\textrm{cosh}(E_{\textrm{eff},m}^f(t) (t-a - T/2))}
\end{eqnarray}
and $\Delta E_{\textrm{eff},m}(t)$ is the statistical error of $E_{\textrm{eff},m}(t)$).

\item $t_\textrm{max}$ is the largest temporal separation $t$, where
\begin{eqnarray}
\Big|E_{\textrm{eff},m}^f(t) - E_{\textrm{eff},m}(t)\Big| \leq 3.5 \times \Delta E_{\textrm{eff},m}(t)
\end{eqnarray}
as well as
\begin{eqnarray}
\frac{\Delta E_{\textrm{eff},m}(t)}{\Delta E_{\textrm{eff},m}(t_\textrm{min})} \leq 3.5 .
\end{eqnarray}
\end{itemize}
This definition of $t_\textrm{min}$ and $t_\textrm{max}$ guarantees that the effective energy is consistent with a plateau within statistical errors for $t \geq t_\textrm{min}$ and that its statistical errors are still reasonably small at $t = t_\textrm{max}$. The energy level $\mathcal{E}_m$ is then determined by averaging $E_{\textrm{eff},m}^f(t)$ over the fitting region,
\begin{eqnarray}
\mathcal{E}_m = \frac{1}{(t_\textrm{max} - t_\textrm{min})/a + 1} \sum_{t=t_\textrm{min}}^{t_\textrm{max}} E_{\textrm{eff},m}^f(t)
\end{eqnarray}
(see also Ref.\ \cite{Donnellan:2010mx}, where a similar procedure was used).
 
The components of the eigenvectors $\mathbf{v}_m(t,t_0)$ obtained by solving the GEVP (\ref{gevp2}) provide information about the structure of the corresponding energy eigenstates:
\begin{equation}
\label{GEPcoefs} | m \rangle \approx \sum_j v^j_m(t,t_0) \mathcal{O}^{j \dagger} | \Omega \rangle ,
\end{equation}
for sufficiently large $t$, where the $\approx$ sign denotes the expansion of the energy eigenstate $| m \rangle$ within the subspace spanned by the trial states $\mathcal{O}^{j \dagger} | \Omega \rangle$. We found that for $t \geq t_\textrm{min}$ the eigenvector components are constant within statistical errors. Thus, we average the eigenvector components $v^j_m(t,t_0)$ according to
\begin{equation}
\label{EQN459} v^j_m = \frac{1}{(t_\textrm{max} - t_\textrm{min}) / a + 1} \sum_{t = t_\textrm{min}}^{t_\textrm{max}} v^j_m(t,t_0)
\end{equation}
and normalize via $v^j_m \rightarrow v^j_m / |\mathbf{v}_m|$.


\subsection{\label{sec:amias}AMIAS method}

In practice, effective energies $E_{\textrm{eff},m}(t)$ often exhibit strong statistical fluctuations, in particular for large $t$ and $m > 0$, rendering a reliable identification of plateaus and extraction of energy levels $\mathcal{E}_m$ difficult. Therefore, in addition to the GEVP method we employ an alternative analysis method called AMIAS \cite{Alexandrou:2008bp,Papanicolas:2012sb,Alexandrou:2014mka}.

In section~\ref{sec:euccor} we have discussed that lattice QCD results for correlation functions $C_{j k}(t)$ (see Eq.\ (\ref{EQN789})) with interpolationg fields $\mathcal{O}^j$, $j = 1,\ldots,7$ (see Eqs.\ (\ref{EQN002}) to (\ref{EQN003})) can be parameterized according to Eq.\ (\ref{eq:cosh_fit}). In the $t$ range we are going to consider, $a \leq t \leq 15 \, a$, and for the energy levels $\mathcal{E}_m$ expected, $\cosh$ and $\sinh$ can be approximated by exponential functions, resulting in fit functions
\begin{eqnarray}
\label{EQN006} C_{j k}^{\textrm{fit}}(t) = 2 \sum_m^\textrm{truncated} c^j_{m,\Omega} c^k_{m,\Omega} e^{-\mathcal{E}_m t} .
\end{eqnarray}
The fit parameters $\mathcal{E}_m$ and $c^j_{m,\Omega}$ are real. In the following they are collectively denoted by $\mathcal{A}_r$.

AMIAS determines a probability distribution function (PDF) $\Pi(\mathcal{A}_r)$ for each fit parameter $\mathcal{A}_r$. The estimates for the values of the fit parameters and their uncertainties are the expectation values and the standard deviations of the corresponding PDFs,
\begin{eqnarray}
 & & \overline{\mathcal{A}}_r = \int d\mathcal{A}_r \, \mathcal{A}_r \Pi(\mathcal{A}_r) \\
 & & \Delta \mathcal{A}_r = \bigg(\int d\mathcal{A}_r \, (\mathcal{A}_r - \overline{\mathcal{A}}_r)^2 \Pi(\mathcal{A}_r)\bigg)^{1/2} .
\end{eqnarray}
AMIAS is able to handle a rather large number of parameters using Monte Carlo techniques, i.e.\ it is suited to study several energy eigenstates, if the lattice QCD results for correlation functions are sufficiently precise.

The PDF for the complete set of fit parameters is defined by
\begin{eqnarray}
\label{EQN004} P(\mathcal{A}_1,\mathcal{A}_2,\ldots) = \frac{1}{N} e^{-\chi^2 / 2}
\end{eqnarray}
with appropriate normalization $N$ and
\begin{eqnarray}
\label{eq:chi2} \chi^2 = \sum_{j,k} \sum_{t=t_{\textrm{min}}}^{t_{\textrm{max}}} \frac{(C_{j k}(t)- C_{j k}^{\textrm{fit}}(t))^2}{(\Delta C_{j k}(t))^2} , 
\end{eqnarray}
which is the well-known $\chi^2$ used in uncorrelated $\chi^2$ minimizing fits. $C_{j k}(t)$ denotes the correlation functions computed using lattice QCD with corresponding statistical errors $\Delta C_{j k}(t)$, while $C_{j,k}^{\textrm{fit}}(t)$ is given by Eq.\ (\ref{EQN006}). In principle one can also use a correlated $\chi^2$. Then, however, one has to estimate a covariance matrix, which requires rather precise data and computations on a large number of gauge link configurations (cf.\ e.g.\ Ref.\ \cite{Michael:1993yj} for a detailed discussion).

To obtain the PDF $\Pi(\mathcal{A}_r)$ for a specific fit parameter $\mathcal{A}_r$, one has to integrate Eq.\ (\ref{EQN004}) over all other parameters. In particular, the probability for the parameter $\mathcal{A}_r$ to be inside the interval $[a,b]$ is
\begin{eqnarray}
\nonumber & & \int_a^b d\mathcal{A}_r \, \Pi(\mathcal{A}_r) = \frac{
          \int_a^b d\mathcal{A}_r \,
          \int_{-\infty}^{+\infty} \prod_{s \ne r} d\mathcal{A}_s \,
          e^{-\chi^2/2}
          }
          {
          \int_{-\infty}^{+\infty} \prod_s d\mathcal{A}_s \,
          e^{-\chi^2/2}
          } . \\
\label{eq:montecarlo} & &
\end{eqnarray}
This multi-dimensional integral can be computed with standard Monte Carlo methods. We use a parallel tempering scheme combined with the Metropolis algorithm as described in detail in Ref.\ \cite{Alexandrou:2014mka}. The parallel tempering scheme prevents the algorithm from getting stuck in a region around a local minimum of $\chi^2$ and guarantees ergodicity of the algorithm. 

While we use $t_\textrm{max} = 15 \, a$ in Eq.\ (\ref{eq:chi2}), we vary in our analyses both $t_\textrm{min}$ and the number of terms in the truncated sum in Eq.\ (\ref{EQN006}), until we find a stable region with no observable change in the PDFs for the low-lying energy eigenstates of interest. For a detailed example see Ref.\ \cite{Alexandrou:2014mka}.

The coefficients $c^j_{\Omega,m} = \langle \Omega | \mathcal{O}^j | m \rangle = \langle m | \mathcal{O}^{j \dagger} | \Omega \rangle$ in the fit function (\ref{EQN006}) are the coefficients of the expansions of the trial states $\mathcal{O}^{j \dagger} | \Omega \rangle$ in terms of the energy eigenstates $| m \rangle$, i.e.\
\begin{eqnarray}
\nonumber \mathcal{O}^{j \dagger} | \Omega \rangle \approx \sum_m^\textrm{truncated} | m \rangle \langle m | \mathcal{O}^{j \dagger} | \Omega \rangle = \sum_m^\textrm{truncated} c^j_{\Omega,m} | m \rangle . \\
\label{EQN630} & &
\end{eqnarray}
More interesting, however, is inverting Eq.\ (\ref{EQN630}) and writing the extracted energy eigenstates in terms of the trial states,
\begin{eqnarray}
\label{amiascoefs} | m \rangle \approx \sum_j \tilde{v}^j_m \mathcal{O}^{j \dagger} | \Omega \rangle .
\end{eqnarray}
One can show that the matrix formed by the coefficients $\tilde{v}^j_m$ is the inverse of the matrix formed by the coefficients $c^j_{\Omega,m}$ up to exponentially small corrections, i.e.\
\begin{eqnarray}
\label{EQN245} \sum_j \tilde{v}^j_m c^j_{\Omega,n} \approx \delta_{m,n} .
\end{eqnarray}
Note that the coefficients $\tilde{v}^j_m$ are equivalent to the eigenvector components $v^j_m$ obtained by solving a GEVP (see Eq.\ (\ref{EQN459})) and, thus, the resuls from the two methods can be compared in a meaningful way, after choosing the same normalization $(\tilde{\mathbf{v}}_m)^2 = 1$.


\section{\label{sec:cormat}Analysis of the correlation matrix and numerical results for the $D_{s0}^\ast(2317)$ meson}


\subsection{\label{SEC455}Extraction of energy levels and amplitudes in a finite volume}

Our goal in this section is to determine the two lowest energy levels in the sector with $D_{s0}^\ast(2317)$ quantum numbers in the finite spatial volume $L^3$ of the lattice. From previous results \cite{Mohler:2013rwa,Lang:2014yfa,Bali:2017pdv} we expect that one of the corresponding energy eigenstates is the lowest $D K$ scattering state, while the other has a somewhat smaller energy and represents the $D_{s0}^\ast(2317)$ meson. A precise determination of these two energy levels is necessary to study the infinite volume limit using L\"uscher's finite volume method in section~\ref{SEC762}.

Moreover, in this section we will also investigate the quark content and arrangement of the $D_{s0}^\ast(2317)$ state by studying the eigenvector components $v_m^j$ and the coefficients $\tilde{v}_m^j$ introduced in section~\ref{SEC467}.


\subsubsection{$D$ and $K$ meson masses and $D K$ threshold}

As a preparatory step we computed the masses of the pseudoscalar mesons $D$ and $K$ within our lattice setup. It is rather straightforward to obtain precise values for $m_K$ and $m_D$ from correlation functions of standard interpolating fields
\begin{eqnarray}
 & & \mathcal{O}^{D} = \sum_{\bf{x}} {\bar c}({\bf x}) \gamma_5 u({\bf x}) \\
 & & \mathcal{O}^{K} = \sum_{\bf{x}} {\bar u}({\bf x}) \gamma_5 s({\bf x}) .
\end{eqnarray}
We find
\begin{eqnarray}
\label{EQN824} & & m_D = 1.8445(9) \, \textrm{GeV} \\
\label{EQN823} & & m_K = 0.5965(4) \, \textrm{GeV}
\end{eqnarray}
using the AMIAS analysis method (extracting the masses from the corresponding effective energies, as explained in the context of the GEVP method in section~\ref{SEC496}, leads to compatible results, however, with somewhat larger statistical errors) \footnote{Here and in the following we convert lattice results, which are obtained in units of the lattice spacing $a$, to $\textrm{GeV}$ or $\textrm{fm}$ by multiplying with appropriate powers of $a = 0.0907 \, \textrm{fm}$. The error on the lattice spacing, $\Delta a = 0.0014 \, \textrm{fm}$ is not taken into account.}. Consequently,
\begin{eqnarray}
\label{EDK} m_D + m_K = 2.4411(10) \, \textrm{GeV} ,
\end{eqnarray}
which is the lowest two-meson threshold in the sector with $D_{s0}^\ast(2317)$ quantum numbers and, thus, plays an important role in the interpretation of further results. Also of interest is the energy of a non-interacting $D K$ pair with one quantum of relative momentum $p_\textrm{min} = 2 \pi / L$,
\begin{eqnarray}
\nonumber & & \Big(m_D^2 + p_\textrm{min}^2\Big)^{1/2} + \Big(m_K^2 + p_\textrm{min}^2\Big)^{1/2} = 2.6271(9) \, \textrm{GeV} . \\
\label{EDK1} & &
\end{eqnarray}



%


\subsubsection{Preselection of interpolating fields}

To reduce the seven interpolating fields (\ref{EQN002}) to (\ref{EQN003}) to a somewhat smaller set of interpolating fields, which are most important to resolve the lowest energy eigenstates with $D_{s0}^\ast(2317)$ quantum numbers, we performed GEVP as well as AMIAS analyses using individual correlation functions or $2 \times 2$ correlation matrices. From these analyses it became clear, that three of the interpolating fields (\ref{EQN002}) to (\ref{EQN003}) are less relevant.
\begin{itemize}
\item $\mathcal{O}^{q \bar{q}, \ \gamma_0}$ (Eq.\ (\ref{EQN002_})): \\
One can determine the energy of a low lying state, which we will identify below as the $D_{s0}^\ast(2317)$ meson, using the correlation function of one of the two quark-antiquark interpolating fields, i.e.\ either of $\mathcal{O}^{q \bar{q}, \ 1}$ or of $\mathcal{O}^{q \bar{q}, \ \gamma_0}$. For the latter, however, the effective energy plateau is reached at larger temporal separation and statistical errors are larger as well. An analysis of the corresponding $2 \times 2$ correlation matrices gives the same low lying state and a second rather noisy effective energy significantly above, which most likely receives contributions from several excited states. The eigenvector components $v_m^j$ indicate a strong dominance of the interpolating field $\mathcal{O}^{q \bar{q}, \ 1}$ for the ground state. In view of these findings we consider $\mathcal{O}^{q \bar{q}, \ 1}$ superior to $\mathcal{O}^{q \bar{q}, \ \gamma_0}$ and do not use the latter interpolating field in any of the following analyses.

\item $\mathcal{O}^{D_s \eta, \ \textrm{point}}$ and $\mathcal{O}^{D_s \eta, \ \textrm{2part}}$ (Eqs.\ (\ref{EQN117}) and (\ref{EQN003})): \\
Correlation functions containing either $\mathcal{O}^{D_s \eta, \ \textrm{point}}$ or $\mathcal{O}^{D_s \eta, \ \textrm{2part}}$ exhibit large statistical errors. This seems to be a consequence of the ``$\eta$ interpolator'' $\bar{u} \gamma_5 u + \bar{d} \gamma_5 d$, which is part of $\mathcal{O}^{D_s \eta, \ \textrm{point}}$ as well as of $\mathcal{O}^{D_s \eta, \ \textrm{2part}}$ (note that a lattice QCD study of the $\eta$ meson is quite challenging by itself, partly because of strong statistical fluctuations; see e.g.\ Refs.\ \cite{Ottnad:2012fv,Michael:2013gka} for a detailed discussion and a recent computation). Moreover, the mass of the $D_{s0}^\ast(2317)$ meson is close to the $D K$ threshold, while the $D_s \eta$ threshold is around $155 \, \textrm{MeV}$ above \cite{Tanabashi:2018oca}. This suggests that including the interpolating fields $\mathcal{O}^{D_s \eta, \ \textrm{point}}$ and $\mathcal{O}^{D_s \eta, \ \textrm{2part}}$ is not essential to resolve the two lowest energy eigenstates. This is supported by Refs.\ \cite{MartinezTorres:2017bdo,Du:2017zvv,Albaladejo:2018mhb,Guo:2018tjx}, where the molecular components for the $D_{s0}^\ast(2317)$ were found to be from around $60 \%$ to $75 \%$ for $D K$ and below $15 \%$ for $D_s \eta$. Thus, we do not use $\mathcal{O}^{D_s \eta, \ \textrm{point}}$ and $\mathcal{O}^{D_s \eta, \ \textrm{2part}}$ in any of the following analyses.



\end{itemize} 
The remaining four interpolating fields $\mathcal{O}^{q \bar{q}, \ 1}$, $\mathcal{O}^{D K, \ \textrm{point}}$, $\mathcal{O}^{Q \bar{Q}, \, \gamma_5}$ and $\mathcal{O}^{D K, \ \textrm{2part}}$ (Eqs.\ (\ref{EQN002}), (\ref{EQN007}), (\ref{EQN631}) and (\ref{EQN113})) are used in the following to determine the two lowest energy levels in the sector with $D_{s0}^\ast(2317)$ quantum numbers.


\subsubsection{\label{SEC508}GEVP analysis}

\begin{figure}[htb]
\begin{center}
\includegraphics[width=8.2cm]{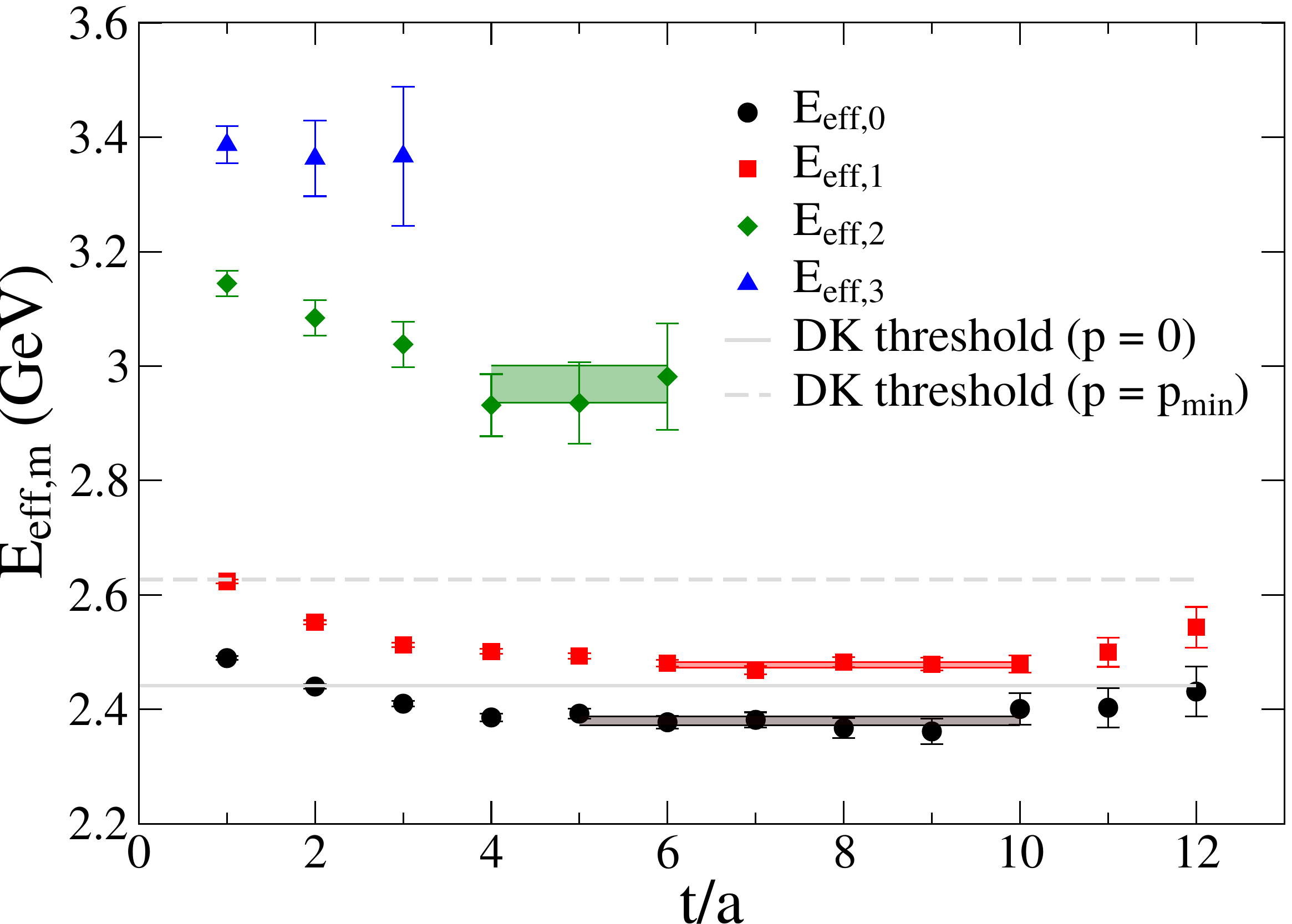} \\
\vspace{0.4cm}
\includegraphics[width=4.1cm,page=1]{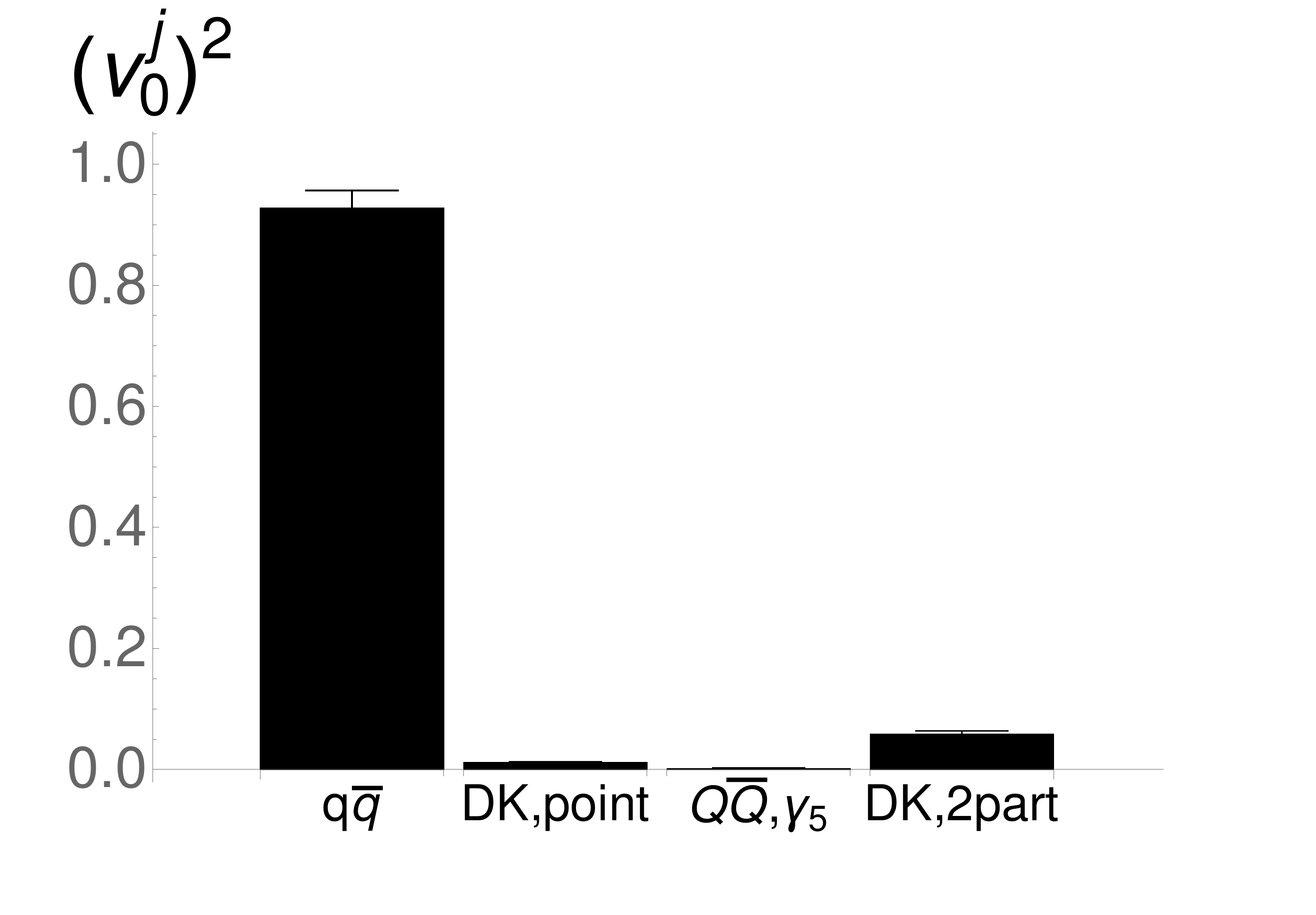}
\includegraphics[width=4.1cm,page=2]{figures/vsquared_GEVP.pdf} \\
\includegraphics[width=4.1cm,page=3]{figures/vsquared_GEVP.pdf}
\includegraphics[width=4.1cm,page=4]{figures/vsquared_GEVP.pdf}
\end{center}
\caption{\label{FIG005}GEVP analysis of the $4 \times 4$ correlation matrix with interpolating fields $\mathcal{O}^{q \bar{q}, \ 1}$, $\mathcal{O}^{D K, \ \textrm{point}}$, $\mathcal{O}^{Q \bar{Q}, \, \gamma_5}$ and $\mathcal{O}^{D K, \ \textrm{2part}}$. (top)~Effective energies $E_{\textrm{eff},m}$ as functions of the temporal separation $t$ together with the $D K$ threshold (see Eq.\ (\ref{EDK})) and the energy of a non-interacting $D K$ pair with one quantum of relative momentum (see Eq.\ (\ref{EDK1})). (bottom)~Squared eigenvector components $(v_m^j)^2$.}
\end{figure}

The results of a GEVP analysis of the $4 \times 4$ correlation matrix containing the four interpolating fields identified in the previous subsection are collected in FIG.\ \ref{FIG005}. The upper plot shows effective energies as functions of the temporal separation. The four plots below contain the squared eigenvector components $(v_m^j)^2$.

There are two convincing effective energy plateaus with small statistical errors close to the $D K$ threshold. One is around $60 \, \textrm{MeV}$ below, while the other is somewhat above, but significantly closer to the $D K$ threshold than to the energy of a non-interacting $D K$ pair with one quantum of relative momentum as indicated by the horizontal gray lines (see also Eqs.\ (\ref{EDK}) and (\ref{EDK1})). Thus, there is an additional low-lying state compared to the non-interacting $D K$ spectrum. The eigenvector components clearly indicate that the lowest state resembles a quark-antiquark pair ($(v_0^{q \bar{q}, \ 1})^2 \gtapprox 0.90$; black bar chart), while the first excitation is a $D K$ scattering state similar to a non-interacting two-meson state with both mesons at rest ($(v_1^{D K, \ \textrm{2part}})^2 \approx 0.85$; red bar chart). These eigenvector components suggest to identify the lowest state as the $D_{s0}^\ast(2317)$ meson. Energy levels $\mathcal{E}_m$ are determined from effective energies as discussed in detail in section~\ref{SEC496}. For $m = 0,1$ they are listed in Table~\ref{TAB640}.


\begin{table}[htb]
\centering
\begin{tabular}{l|c|c}
\hline
 & & \vspace{-0.35cm} \\
analysis & $\mathcal{E}_0 / \textrm{GeV}$ & $\mathcal{E}_1 / \textrm{GeV}$ \\
 & & \vspace{-0.35cm} \\
\hline
GEVP, $4 \times 4$                & $2.3803(78)$ & $2.4780(50)\phantom{0}$ \\
\hline
AMIAS, $4 \times 4$               & $2.3790(28)$ & $2.4854(44)\phantom{0}$ \\
AMIAS, $3 \times 3$, (A) & $2.3765(33)$ & $2.4837(49)\phantom{0}$ \\ 
AMIAS, $3 \times 3$, (B) & $2.3794(35)$ & $2.4946(36)\phantom{0}$ \\ 
AMIAS, $3 \times 3$, (C) & $2.3857(94)$ & $2.5028(135)$           \\ 
AMIAS, $3 \times 3$, (D) & $2.3953(69)$ & $2.7840(456)$           \\ 
\hline
\end{tabular}
\caption{\label{TAB640}The lowest two energy levels $\mathcal{E}_0$ and $\mathcal{E}_1$ in the sector with $D_{s0}^\ast(2317)$ quantum numbers in the finite volume $L^3$ of the lattice obtained by various analyses. (A), (B), (C) and (D) refer to the $3 \times 3$ AMIAS analyses discussed in section~\ref{SEC965}.}
\end{table}


The third effective energy $E_{\textrm{eff},2}(t)$ has large statistical errors and is somewhat above the estimated energy of a non-interacting $D K$ pair with one quantum of relative momentum. It seems likely that it corresponds to a linear superposition of several $D K$ scattering states with non-vanishing relative momenta. This interpretation is supported by the eigenvector components, which indicate a dominance of the $\mathcal{O}^{D K, \ \textrm{point}}$ interpolating field ($(v_2^{D K, \ \textrm{point}})^2 \approx 0.90$; green bar chart), which by construction excites $D K$ states with many different relative momenta. The fourth effective energy $E_{\textrm{eff},2}(t)$ has even larger statistical errors and is around $1 \, \textrm{GeV}$ above the $D K$ threshold. Most likely it represents a superposition of a larger number of highly excited states.

From $(v_m^{Q \bar{Q}, \, \gamma_5})^2 \ltapprox 0.05$ for $m = 0, 1, 2$ one can conclude that the diquark-antidiquark interpolating field $\mathcal{O}^{Q \bar{Q}, \, \gamma_5}$ is not important to resolve any of the three lowest energy eigenstates. In particular the ground state seems to be predominantly a standard quark-antiquark pair ($(v_0^{q \bar{q}, \ 1})^2 \gtapprox 0.90$) with only a small $D K$ component ($(v_0^{D K, \ \textrm{2part}})^2 \ltapprox 0.10$). There is no significant contribution from the tetraquark interpolating fields, i.e.\ both $(v_0^{D K, \ \textrm{point}})^2$ and $(v_0^{Q \bar{Q}, \, \gamma_5})^2$ are almost vanishing. We interpret this as indication that the $D_{s0}^\ast(2317)$ meson has no sizable tetraquark component.


\subsubsection{\label{SEC965}AMIAS analysis}

\begin{figure}[htb]
\begin{center}
\includegraphics[width=8.2cm]{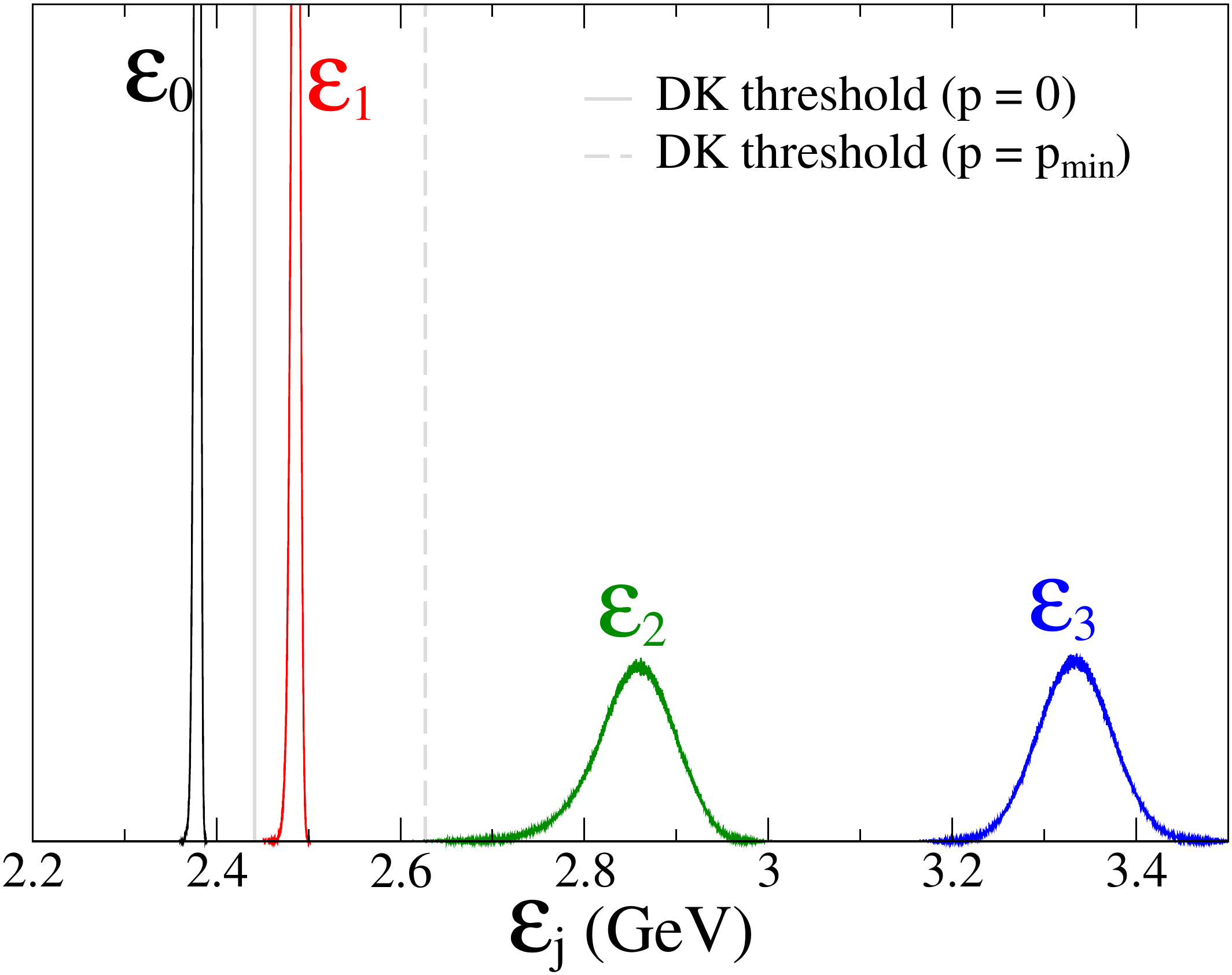} \\
\vspace{0.4cm}
\includegraphics[width=4.1cm,page=1]{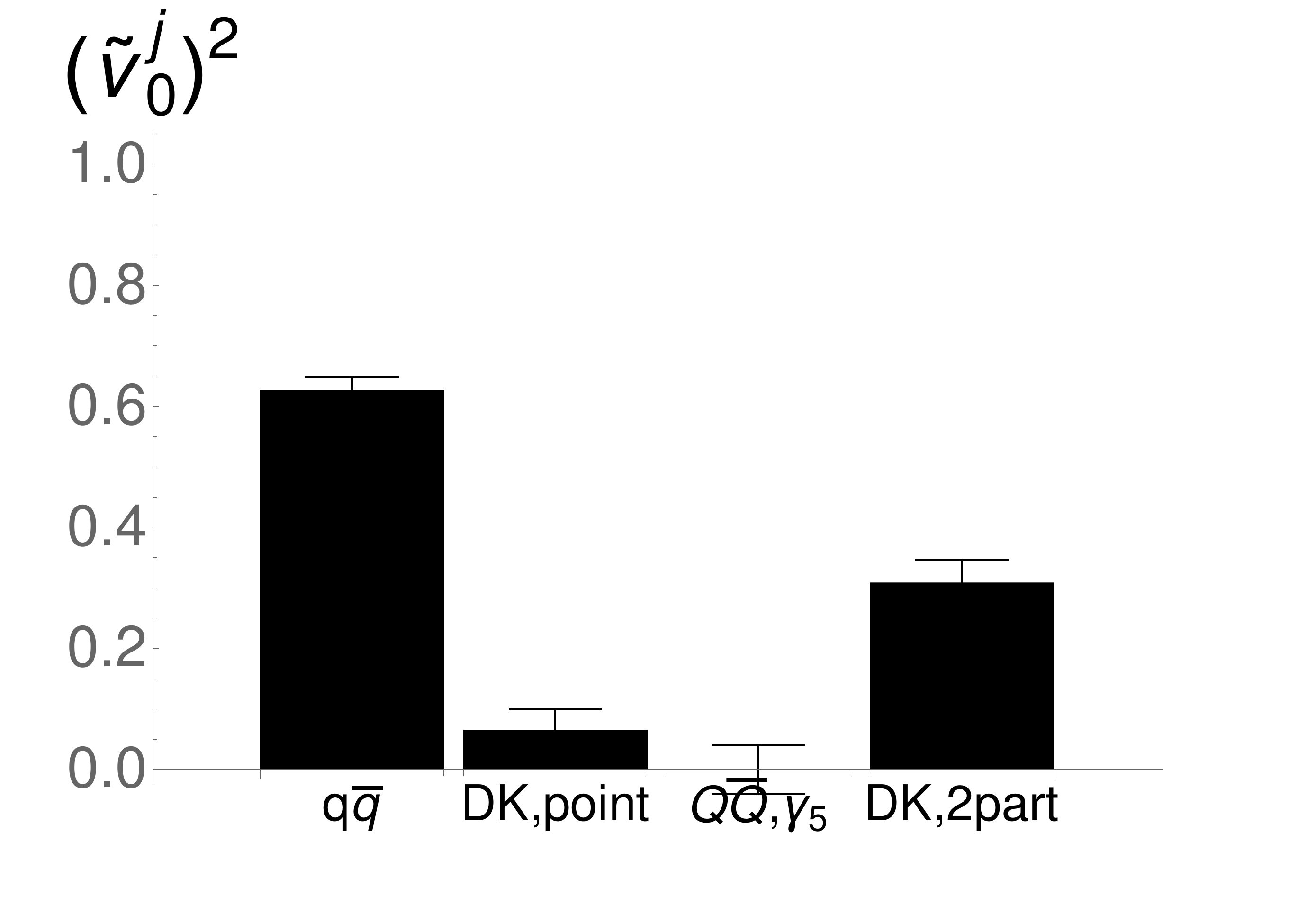}
\includegraphics[width=4.1cm,page=2]{figures/vsquared_AMIAS.pdf} \\
\includegraphics[width=4.1cm,page=3]{figures/vsquared_AMIAS.pdf}
\includegraphics[width=4.1cm,page=4]{figures/vsquared_AMIAS.pdf}
\end{center}
\caption{\label{FIG055}AMIAS analysis of the $4 \times 4$ correlation matrix with interpolating fields $\mathcal{O}^{q \bar{q}, \ 1}$, $\mathcal{O}^{D K, \ \textrm{point}}$, $\mathcal{O}^{Q \bar{Q}, \, \gamma_5}$ and $\mathcal{O}^{D K, \ \textrm{2part}}$. (top)~PDFs for the energy levels together with the $D K$ threshold (see Eq.\ (\ref{EDK})) and the energy of a non-interacting $D K$ pair with one quantum of relative momentum (see Eq.\ (\ref{EDK1})). (bottom)~Squared coefficients $(\tilde{v}^j_m)^2$ for the four lowest energy levels.}
\end{figure}

The results of an AMIAS analysis of the $4 \times 4$ correlation matrix are collected in FIG.\ \ref{FIG055}. The upper plot shows the PDFs generated with the fit function given in Eq.\ (\ref{EQN006}) and six terms in the truncated sum \footnote{Six terms in the truncated sum led to stable results for the energy differences $\mathcal{E}_0$ to $\mathcal{E}_5$, i.e.\ there is no significant change in the corresponding PDFs, when using more than six terms.}. The four plots below show the squared coefficients $(\tilde{v}_m^j)^2$.

When comparing the PDFs to the effective energies in FIG.\ \ref{FIG005} one can see, that the energy levels obtained with AMIAS are consistent with those from the GEVP analysis. Statistical errors for the AMIAS results are somewhat smaller than for the GEVP results (see Table~\ref{TAB640}). The coefficients $(\tilde{v}_m^j)^2$ are also in reasonable agreement with the GEVP eigenvector components $(v_m^j)^2$ from FIG.\ \ref{FIG005}, supporting that the ground state is mostly a quark antiquark-pair.

\begin{figure}[htb]
\begin{center}
\includegraphics[width=8.2cm]{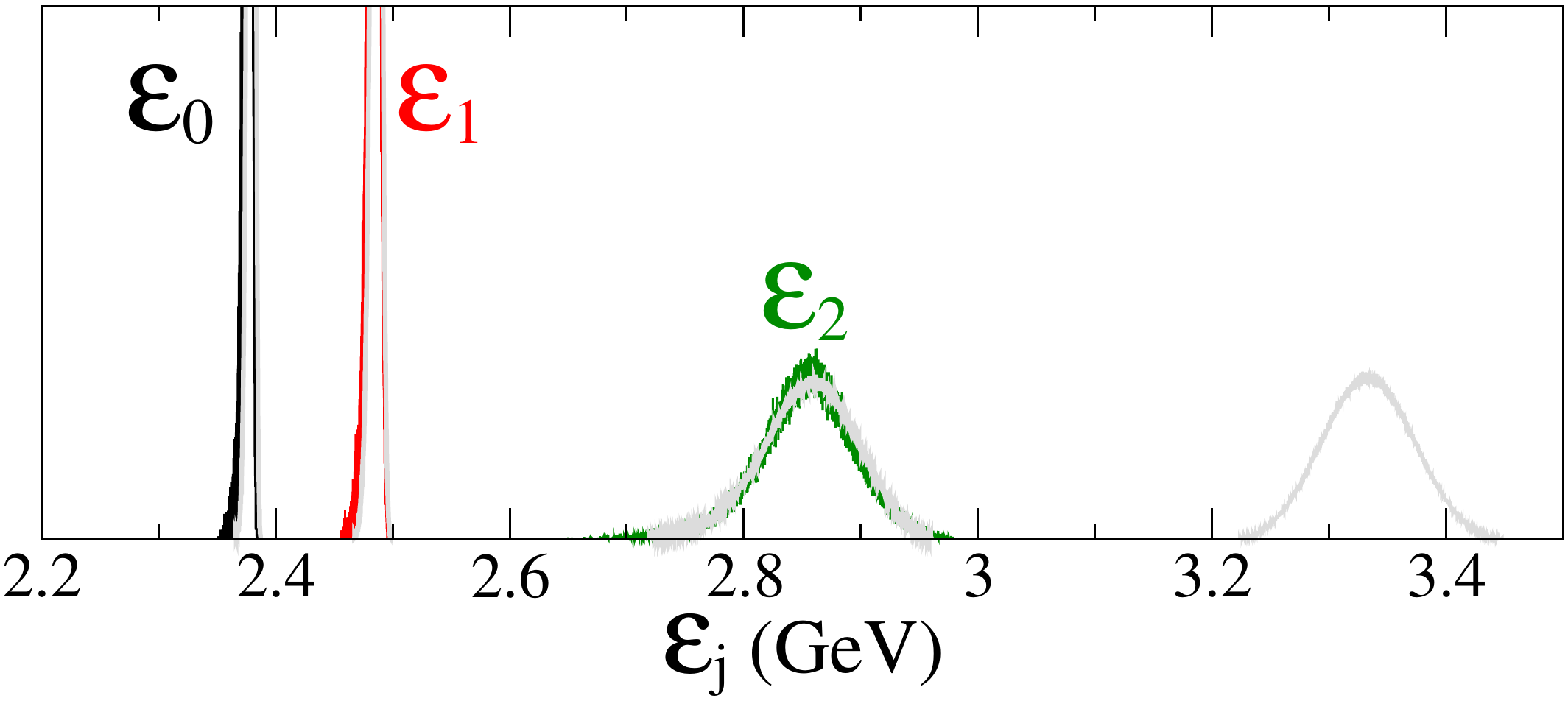} \\  
\includegraphics[width=8.2cm]{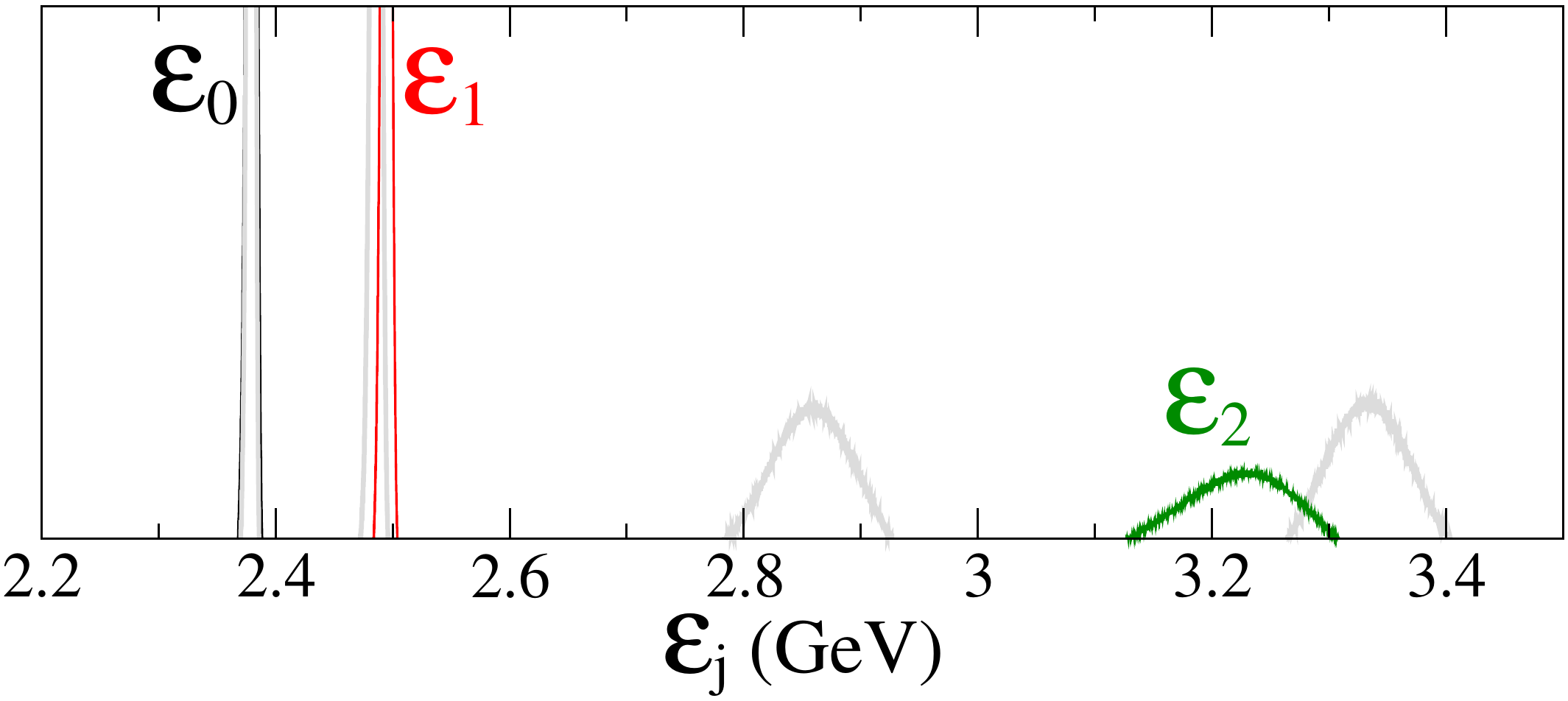} \\  
\includegraphics[width=8.2cm]{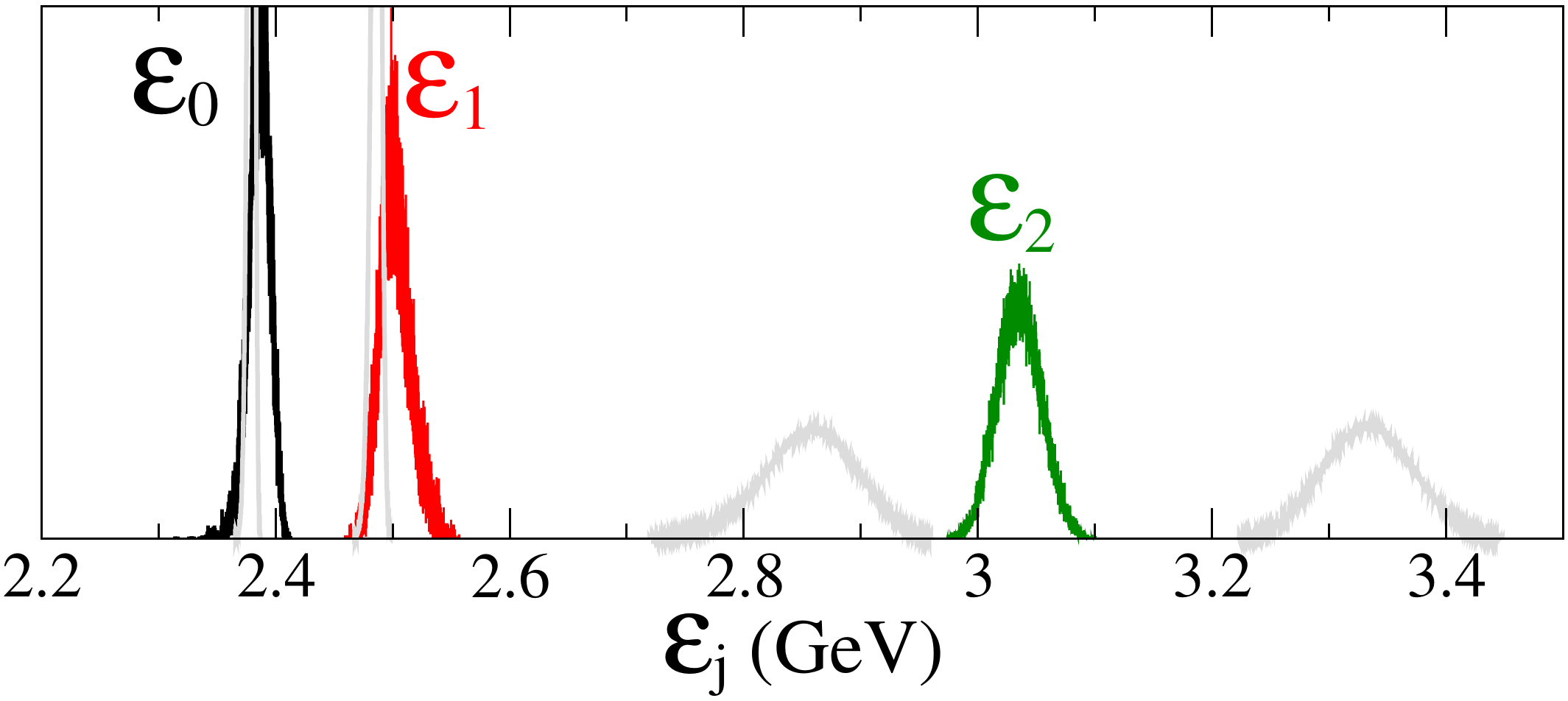} \\  
\includegraphics[width=8.2cm]{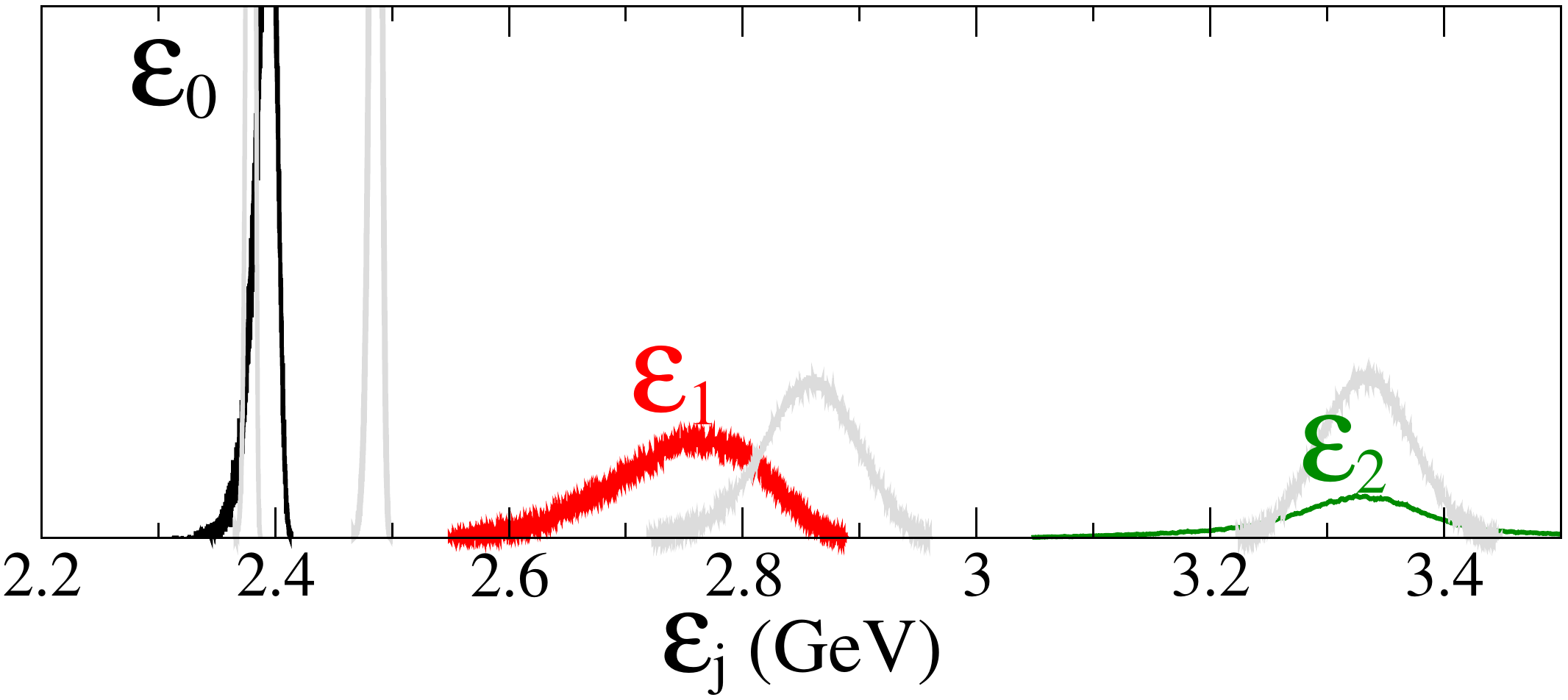}     
\end{center}
\caption{\label{FIG003}PDFs for the energy levels from $3 \times 3$ AMIAS analyses. (A), (B), (C) and (D) refer to the $3 \times 3$ correlation matrices discussed in section~\ref{SEC965}. The light gray PDFs in the background correspond to the $4 \times 4$ AMIAS analysis and are shown to facilitate comparison.}
\end{figure}

To cross-check the obtained results, in particular to confirm our findings regarding the quark composition and interpretation of the low-lying energy eigenstates, it is useful to compare the above $4 \times 4$ AMIAS analysis to analogous analyses using the the four possible $3 \times 3$ submatrices as input (for the latter five terms in the truncated sum in Eq.\ (\ref{EQN006}) are sufficient). The corresponding PDFs are shown in FIG.\ \ref{FIG003}, with the $4 \times 4$ PDFs in the background colored in light gray.
\begin{itemize}
\item[(A)] $3 \times 3$ \textbf{correlation matrix without} $\mathcal{O}^{Q \bar{Q}, \, \gamma_5}$:
\\ There is essentially no difference between the $3 \times 3$ and $4 \times 4$ PDFs for the three lowest energy levels. This confirms that the diquark-antidiquark interpolating field $\mathcal{O}^{Q \bar{Q}, \, \gamma_5}$ is not important to resolve the low-lying energy eigenstates. This in turn supports our conclusion from section~\ref{SEC508} that the $D_{s0}^\ast(2317)$ meson does not have a sizable tetraquark component.

\item[(B)] $3 \times 3$ \textbf{correlation matrix without} $\mathcal{O}^{D K, \ \textrm{point}}$:
\\ The lowest two energy levels are consistent with the $4 \times 4$ result within statistical errors. The energy level of the second excitation is, however, significantly larger. This indicates that the interpolating field $\mathcal{O}^{D K, \ \textrm{point}}$ is useful to resolve higher momentum excitations, while it is not essential for a determinaton of the lowest two energy levels.

\item[(C)] $3 \times 3$ \textbf{correlation matrix without} $\mathcal{O}^{q \bar{q}, \ 1}$:
\\ The lowest two energy levels are slightly larger compared to the $4 \times 4$ result, but they are still compatible, because of their drastically larger statistical errors (see Table~\ref{TAB640}). Thus, it is possible to excite the $D_{s0}^\ast(2317)$ meson with only four-quark interpolating fields, i.e.\ it seems to have a non-vanishing, but small $D K$ component, which is in agreement with our findings from section~\ref{SEC508}.

\item[(D)] $3 \times 3$ \textbf{correlation matrix without} $\mathcal{O}^{D K, \ \textrm{2part}}$:
\\ The lowest energy level is consistent with the $4 \times 4$ result, even though it has a much larger statistical error. The energy level of the first excitation, however, cannot be determined reliably anymore. The PDF has a large width and its peak is localized at energies significantly above the $D K$ threshold. This confirms that the interpolating field $\mathcal{O}^{D K, \ \textrm{2part}}$ is of central importance for a determination the energy of the lowest $D K$ scattering state.
\end{itemize}

\subsubsection{Summary of finite volume results and conclusions}

\begin{figure}[htb]
\begin{center}
\includegraphics[width=8.2cm]{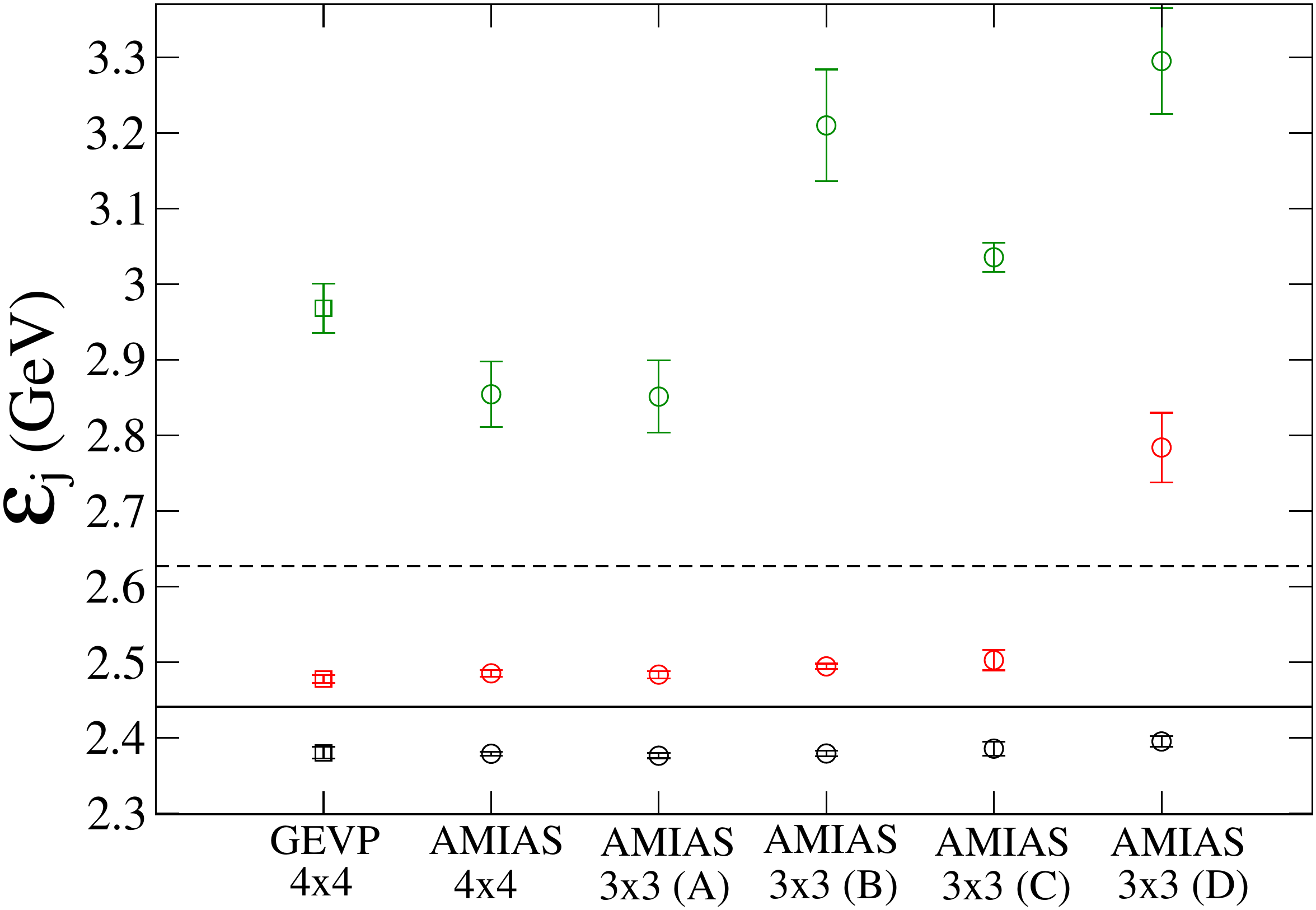}
\end{center}
\caption{\label{FIG004}Comparison plot of finite volume energy levels obtained from $3 \times 3$ and $4 \times 4$ correlation matrices using the GEVP method and the AMIAS method. (A), (B), (C) and (D) refer to the $3 \times 3$ AMIAS analyses discussed in section~\ref{SEC965}.}
\end{figure}

A summary plot of the obtained energy levels with the $4 \times 4$ GEVP as well as the $4 \times 4$ and $3 \times 3$ AMIAS analyses is shown in FIG.\ \ref{FIG004}. Again it can be seen that the most important interpolating fields to determine the two lowest energy levels are $\mathcal{O}^{q \bar{q}, \ 1}$ and $\mathcal{O}^{D K, \ \textrm{2part}}$. Analyses using these two interpolating fields ($4 \times 4$ GEVP, $4 \times 4$ AMIAS, $3 \times 3$ AMIAS (A) and (B)) yield consistent energy levels with small statistical errors.

From the GEVP eigenvector components $v^j_m$ and the AMIAS coefficients $\tilde{v}^j_m$ we conclude that the lowest energy level mostly corresponds to a quark-antiquark bound state, possibly similar to the $D_{s0}^\ast(2317)$ meson in the infinite volume. There seems to be a small $D K$ component, but no sign of any sizable tetraquark component. Moreover, the components $v^j_m$ and the coefficients $\tilde{v}^j_m$ clearly indicate that the first excitation is a $D K$ scattering state. These two energy levels will be used for the finite volume analysis in section~\ref{SEC762}.


\subsection{\label{SEC762}Scattering analysis and infinite volume limit}

The energy levels collected in Table~\ref{TAB640} were computed at finite spatial volume $L^3$ with periodic boundary conditions. One can determine the mass of the $D_{s0}^\ast(2317)$ meson, which is the infinite volume limit of the ground state energy, from the lowest two energy levels $\mathcal{E}_0$ and $\mathcal{E}_1$ at finite volume by performing a scattering analysis continued to imaginary momenta, i.e.\ using L\"uscher's finite volume method \cite{Luscher:1990ux}. This approach has been used in a lattice QCD study of the $D_{s0}^\ast(2317)$ meson for the first time in Refs.\ \cite{Mohler:2013rwa,Lang:2014yfa} and later also in Ref.\ \cite{Bali:2017pdv}. It was also used to study other systems (see e.g.\ Refs.\ \cite{Moir:2016srx,Leskovec:2019ioa}). For a recent review on scattering in lattice QCD see Ref.\ \cite{Briceno:2017max}.

The first step is to determine the squared scattering momenta $k_0^2$ and $k_1^2$ via
\begin{eqnarray}
\label{EQN402} \mathcal{E}_n = \Big(m_D^2 + k_n^2\Big)^{1/2} + \Big(m_K^2 + k_n^2\Big)^{1/2} ,
\end{eqnarray}
where $\mathcal{E}_0$ and $\mathcal{E}_1$ can be taken from Table~\ref{TAB640} and $m_D$ and $m_K$ are the $D$ meson and $K$ meson masses obtained within the same lattice setup (see Eqs.\ (\ref{EQN824}) and (\ref{EQN823})). With L\"uschers finite volume method one can then compute the corresponding two phase shifts $\delta_0(k_0)$ and $\delta_0(k_1)$,
\begin{eqnarray}
\label{EQN679} k_n \cot(\delta_0(k_n)) = \frac{2 \mathcal{Z}_{00}(1;(k_n L/2 \pi)^2)}{\sqrt{\pi} L} .
\end{eqnarray}
Here $\mathcal{Z}_{00}$ denotes the generalized zeta function and $L \approx 2.90 \, \textrm{fm}$ the spatial lattice extent (see section~\ref{sec:simu}).

$k \cot(\delta_0(k))$ can be written as a Taylor series in $k^2$,
\begin{eqnarray}
\label{EQN401} k \cot(\delta_0(k)) = \frac{1}{a_0} + \frac{r_0}{2} k^2 + \mathcal{O}(k^4) ,
\end{eqnarray}
where $a_0$ is the $S$ wave scattering length and $r_0$ the $S$ wave effective range. For sufficiently small $k^2$ one can neglect terms of order $k^4$ and parameterize $k \cot(\delta_0(k))$ by the first two terms on the right hand side of Eq.\ (\ref{EQN401}). $a_0$ and $r_0$ are then fixed by the two data points $\cot(\delta_0(k_0))$ and $\cot(\delta_0(k_1))$ obtained via Eq.\ (\ref{EQN679}). This parameterization is called effective range expansion.

A stable $D_{s0}^\ast(2317)$ meson manifests itself as a pole in the scattering amplitude,
\begin{eqnarray}
\label{EQN403} f_0 k = \frac{1}{\cot(\delta_0(k)) - i} ,
\end{eqnarray}
i.e.\ corresponds to $\cot(\delta_0(k_{D_{s0}^\ast})) = i$, where $k_{D_{s0}^\ast}$ denotes the position of the pole. Combining this condition with the parameterization (\ref{EQN401}) leads to
\begin{eqnarray}
\label{EQN046} i k_{D_{s0}^\ast} = \frac{1}{a_0} + \frac{r_0}{2} k_{D_{s0}^\ast}^2 ,
\end{eqnarray}
which can easily be solved with respect to $k_{D_{s0}^\ast}^2$,
\begin{eqnarray}
k_{D_{s0}^\ast}^2 = -\bigg(\frac{1}{r_0} \pm \bigg(\frac{1}{r_0^2} + \frac{2}{a_0 r_0}\bigg)^{1/2}\bigg)^2
\end{eqnarray}
(note that for our data one of the two solutions has to be discarded, because it is far outside the region of validity of the effective range expansion (\ref{EQN401}), where $\mathcal{O}(k^4)$ terms cannot be neglected). The mass of the $D_{s0}^\ast(2317)$ meson is given by the right hand side of Eq.\ (\ref{EQN402}) with $k_n^2$ replaced by $k_{D_{s0}^\ast}^2$, i.e.\ by
\begin{eqnarray}
m_{D_{s0}^\ast} = \Big(m_D^2 + k_{D_{s0}^\ast}^2\Big)^{1/2} + \Big(m_K^2 + k_{D_{s0}^\ast}^2\Big)^{1/2} .
\end{eqnarray}

In Table~\ref{TAB401} we show the results obtained for the lowest two energy levels $\mathcal{E}_0$ and $\mathcal{E}_1$, for the squared scattering momenta $k_0^2$ and $k_1^2$, for the phase shifts, for the $S$ wave scattering length $a_0$ and effective range $r_0$ as well as for the position of the pole. To verify that the effective range expansion (\ref{EQN401}) is a reasonable approximation, we also provide $s(k_0^2)$, $s(k_1^2)$ and $s(k_{D_{s0}^\ast}^2)$, where $s(k^2) = |a_0 r_0 k^2 / 2|$ corresponds to the ratio of the $\mathcal{O}(k^2)$ term and the $\mathcal{O}(k^0)$ term in Eq.\ (\ref{EQN401}). We find values $\ll 1$ for the two scattering momenta as well as for the position of the pole, which gives certain indication that higher order terms are suppressed, i.e.\ that $\mathcal{O}(k^4)$ terms in Eq.\ (\ref{EQN401}) are indeed negligible. In Table~\ref{TAB401} we also list $m_{D_{s0}^\ast}$, the resulting mass of the $D_{s0}^\ast$ meson, and $m_D + m_K - m_{D_{s0}^\ast}$, the binding energy with respect to the $D K$ threshold. All results are provided both for the $4 \times 4$ GEVP and the $4 \times 4$ AMIAS determination of the lowest two energy levels $\mathcal{E}_0$ and $\mathcal{E}_1$ discussed in sections \ref{SEC508} and \ref{SEC965}.


\begin{table*}[htb]
\centering

\begin{tabular}{l|cc|cc|cc}
\hline
 & & & & & & \vspace{-0.35cm} \\
 & $\mathcal{E}_0 / \textrm{GeV}$ & $\mathcal{E}_1 / \textrm{GeV}$ & $k_0^2 / \textrm{GeV}^2$ & $k_1^2 / \textrm{GeV}^2$ & $k_0 \cot(\delta_0(k_0)) / \textrm{GeV}$ & $k_1 \cot(\delta_0(k_1)) / \textrm{GeV}$ \\
 & & & & & & \vspace{-0.35cm} \\
\hline
GEVP, $4 \times 4$  & $2.3803(78)$ & $2.4780(50)$ & $-0.0531(66)$ & $+0.0340(46)$ & $-0.2101(195)$           & $-0.2350(272)$ \\
AMIAS, $4 \times 4$ & $2.3790(28)$ & $2.4854(44)$ & $-0.0542(25)$ & $+0.0408(38)$ & $-0.2133(72)\phantom{0}$ & $-0.1984(187)$ \\
\hline
\end{tabular}

\vspace{0.2cm}

\begin{tabular}{l|cc|c|ccc|cc}
\hline
 & & & & & & & \vspace{-0.35cm} \\
 & $a_0 / \textrm{fm}$ & $r_0 / \textrm{fm}$ & $k_{D_{s0}^\ast}^2 / \textrm{GeV}^2$ & $s(k_0^2)$ & $s(k_1^2)$ & $s(k_{D_{s0}^\ast}^2)$ & $m_{D_{s0}^\ast} / \textrm{GeV}$ & $(m_D + m_K - m_{D_{s0}^\ast}) / \textrm{MeV}$ \\
 & & & & & & & \vspace{-0.35cm} \\
\hline
GEVP, $4 \times 4$  & $-0.876(76)$ & $-0.113(152)$           & $-0.0451(99)$ &
$0.07$ & $0.04$ & $0.06$ & $2.3897(116)$           & $51.3(11.7)$           \\

AMIAS, $4 \times 4$ & $-0.964(34)$ & $+0.062(84)\phantom{0}$ & $-0.0449(27)$ &
$0.04$ & $0.03$ & $0.03$ & $2.3900(64)\phantom{0}$ & $51.1(6.5)\phantom{0}$ \\
\hline
\end{tabular}

\caption{\label{TAB401}Results of the scattering analysis.}

\end{table*}

\begin{figure}[htb]
\begin{center}
\includegraphics[width=8.2cm]{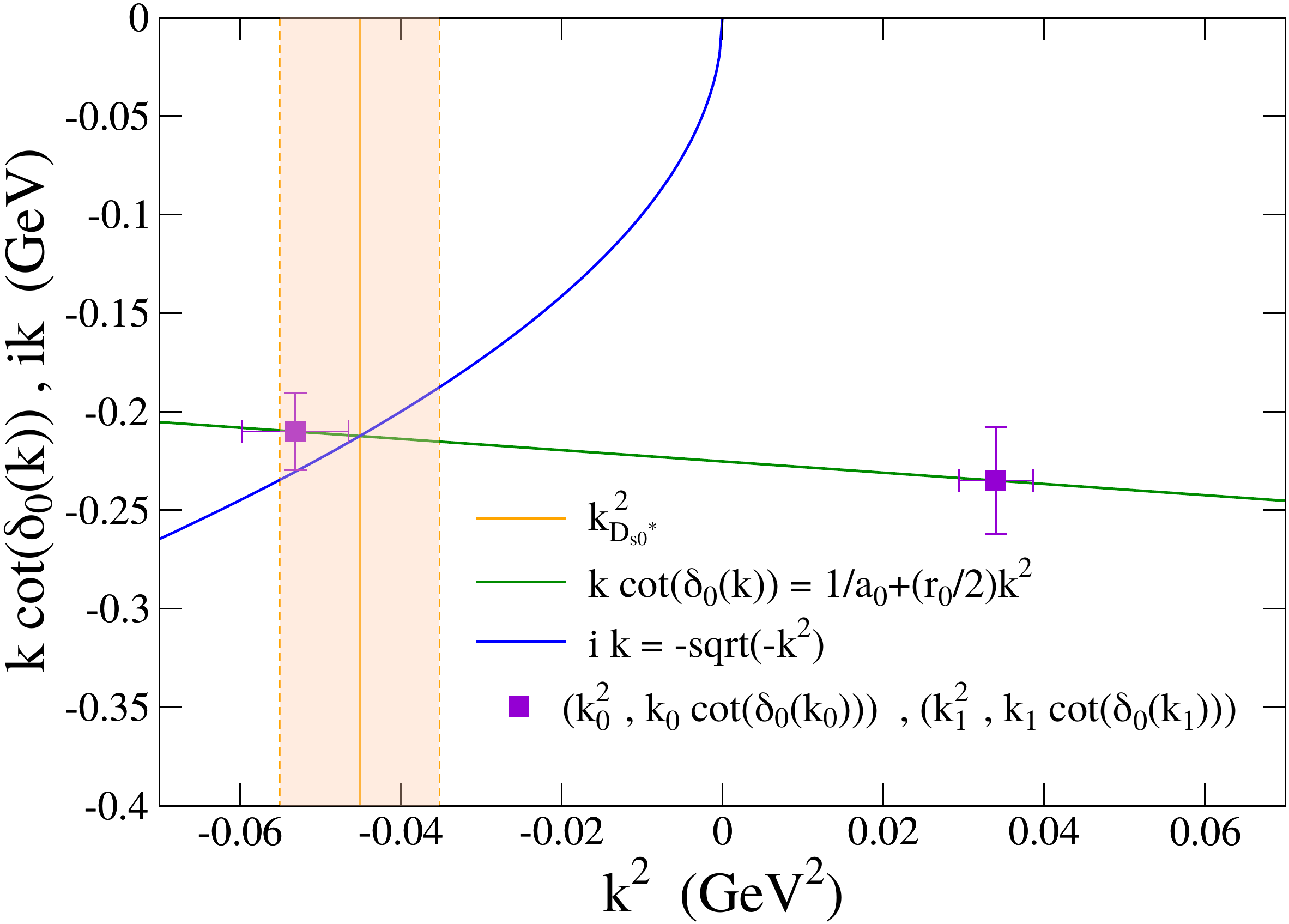} \\
\vspace{0.3cm}
\includegraphics[width=8.2cm]{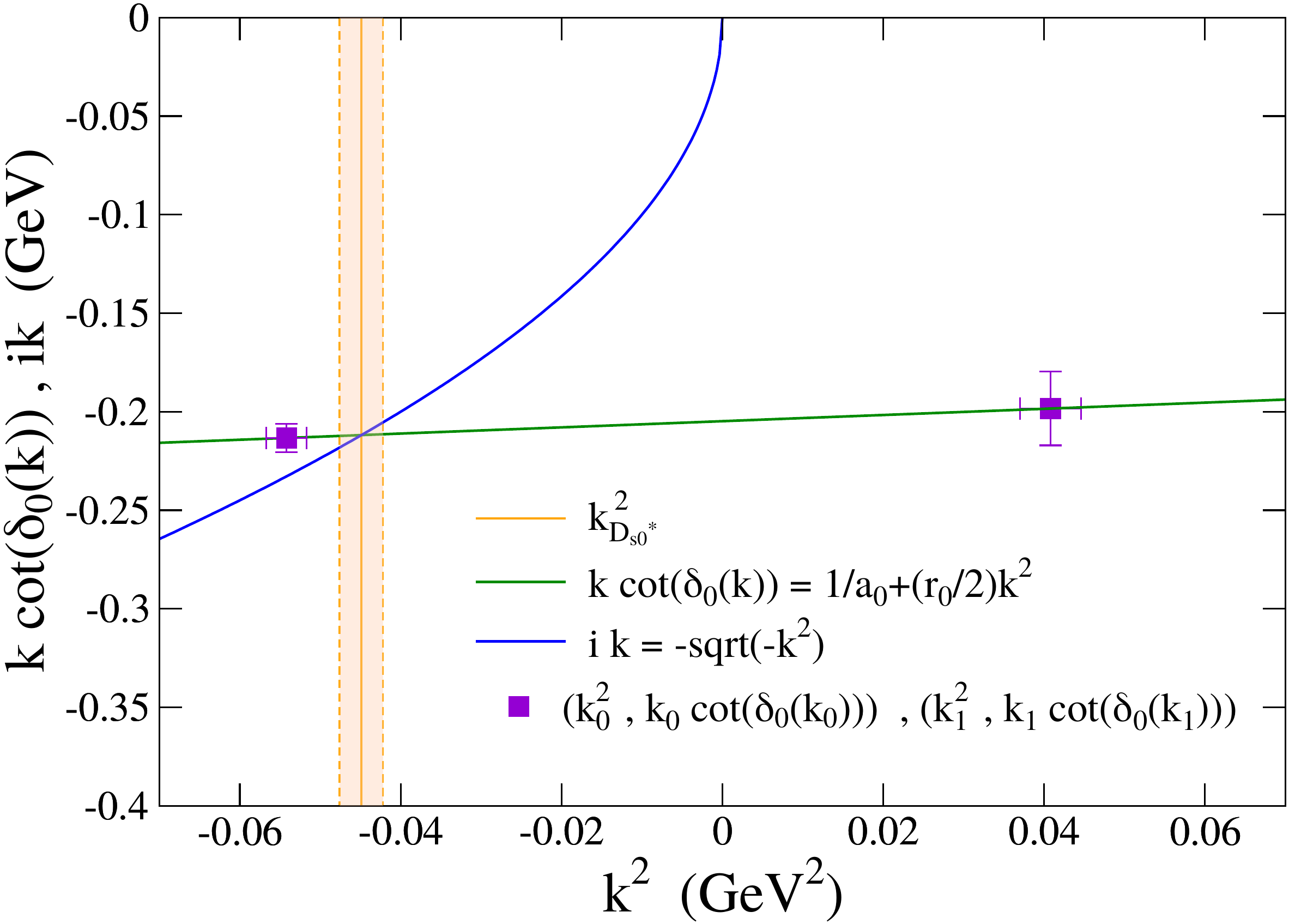}
\end{center}
\caption{\label{FIG401}The parameterization of $k \cot(\delta_0(k))$ with the effective range expansion (right hand side of both Eq.\ (\ref{EQN401}) and Eq.\ (\ref{EQN046}); green curve) together with the two data points $k_0 \cot(\delta_0(k_0))$ and $k_1 \cot(\delta_0(k_1))$ (in magenta). The intersection with the left hand side of Eq.\ (\ref{EQN046}), $i k = -\sqrt{-k^2}$ (blue curve), corresponds to $k_{D_{s0}^\ast}^2$ (indicated by the orange error band). (top)~$4 \times 4$ GEVP analysis. (bottom)~$4 \times 4$ AMIAS analysis.}
\end{figure}

In FIG.~\ref{FIG401} we show the parameterization of $k \cot(\delta_0(k))$ with the effective range expansion (\ref{EQN401}) together with the two data points $k_0 \cot(\delta_0(k_0))$ and $k_1 \cot(\delta_0(k_1))$. Note that the effective range expansion is equivalent to the right hand side of Eq.\ (\ref{EQN046}). We also show the left hand side of that equation, $i k = -\sqrt{-k^2}$. The intersection of the two curves corresponds to $k_{D_{s0}^\ast}^2$, the binding momentum of the $D_{s0}^\ast(2317)$ meson.

\begin{figure}[htb]
\begin{center}
\includegraphics[width=8.2cm]{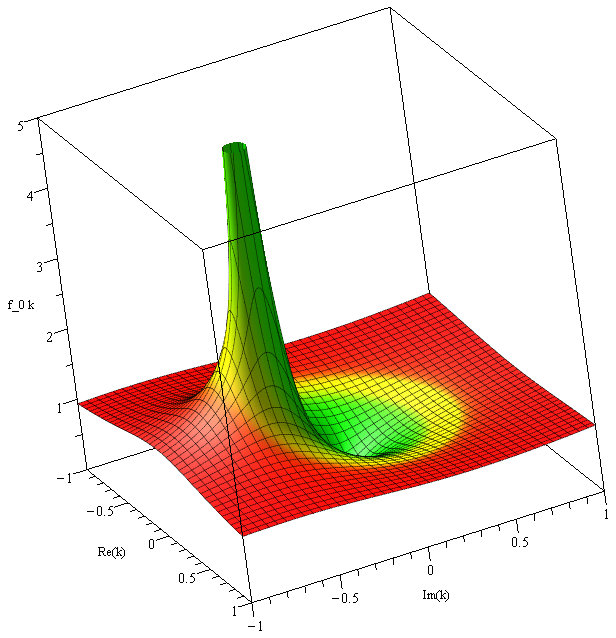}
\end{center}
\caption{\label{FIG402}$|f_0 k|$ according to Eq.\ (\ref{EQN256}) in the complex $k$ plane. The pole corresponds to the $D_{s0}^\ast(2317)$ meson. The color of the plotted surface is related to the value of $s(k^2)$ (green: $s(k^2) < 0.1$; yellow: $0.1 \leq s(k^2) < 0.2$; red: $0.2 \leq s(k^2)$). Thus, it reflects the quality of the effective range expansion (\ref{EQN401}) and indicates that the pole at $k_{D_{s0}^\ast}$ is in a region, where $\mathcal{O}(k^4)$ terms should be negligible.}
\end{figure}

In FIG.~\ref{FIG402} we illustrate the pole in the scattering amplitude by plotting
\begin{eqnarray}
\label{EQN256} |f_0 k| = \bigg|\frac{1}{a_0 k} + \frac{r_0 k}{2} - i\bigg|^{-1}
\end{eqnarray}
in the complex $k$ plane, i.e.\ Eq.\ (\ref{EQN403}) with the parameterization (\ref{EQN401}) inserted. The color reflects the quality of the effective range expansion (\ref{EQN401}) and indicates that the pole is in a region, where $\mathcal{O}(k^4)$ terms should be negligible.

It is important to note that a direct comparison of our result for $m_{D_{s0}^\ast}$ to the corresponding experimental result $m_{D_{s0,\textrm{exp}}^\ast} = 2317.8(5) \, \textrm{MeV}$ \cite{Tanabashi:2018oca} is not meaningful, because the quark masses in our simulation differ from their experimental counterparts:
\begin{itemize}
\item The light $u$ and $d$ quark mass is unphysically heavy, reflected by the pion mass $m_\pi \approx 0.296 \, \textrm{GeV}$.

\item The $s$ quark mass is unphysically heavy, as indicated by $2 m_K^2 - m_\pi^2 \approx 0.62 \, \textrm{GeV}^2$ (which is approximately proportional to the $s$ quark mass) compared to $2 m_{K,\textrm{exp}}^2 - m_{\pi,\textrm{exp}}^2 \approx 0.47 \, \textrm{GeV}^2$.

\item The $c$ quark is unphysically light, because $m_D \approx 1.845 \, \textrm{GeV}$, i.e.\ below $m_{D,\textrm{exp}} \approx 1.867 \, \textrm{GeV}$.
\end{itemize}
Also for the binding energy of the $D_{s0}^\ast(2317)$ meson with respect to the $D K$ threshold, $m_D + m_K - m_{D_{s0}^\ast}$, it is not clear a priori, whether quark masses, which differ from their physical values, will result in a value similar to the corresponding experimental value $m_{D,\textrm{exp}} + m_{K,\textrm{exp}} - m_{D_{s0,\textrm{exp}}}^\ast \approx 45 \, \textrm{MeV}$. One reason for this is that the threshold $m_D + m_K$ clearly depends on the light quark mass, while $m_{D_{s0}^\ast}$, which according to section~\ref{SEC455} is mostly a $\bar{c} s$ state, should be almost independent of the light quark mass (see also the discussion in Ref.\ \cite{Lang:2014yfa}). Note, however, that we find $m_D + m_K - m_{D_{s0}^\ast} \approx 51 \, \textrm{MeV}$ rather close to the experimentally observed $45 \, \textrm{MeV}$, which indicates that with respect to the $D_{s0}^\ast$ meson we might be in a similar situation as in real world QCD, even though we are not precisely at physical quark masses. Because of this and since $m_{D_{s0}^\ast}$ is close to the lowest energy level $\mathcal{E}_0$ obtained at finite lattice volume (around $10 \, \textrm{MeV}$ difference as can be seen from Table~\ref{TAB401}), we expect that our findings and statements from section~\ref{sec:cormat} about the importance of the two-quark and the four quark interpolating fields also apply for physical quark masses and the infinite volume limit. This is further supported by the qualitative agreement of our results for $a_0$ and $r_0$ and corresponding results obtained in lattice QCD computations at almost physical quark masses \cite{Lang:2014yfa,Bali:2017pdv}.


\section{\label{sec:conclusions}Summary and conclusions}

We studied the $D_{s0}^\ast(2317)$ meson with lattice QCD using interpolating fields of different structure. In addition to quark-antiquark interpolating fields and two-meson interpolating fields, which were already considered in previous lattice QCD studies, we implemented and explored the importance of tetraquark interpolating fields. For these tetraquark interpolating fields the four quark operators are centered at the same point in space and their color and spin structure corresponds to either a meson-meson pair or a diquark-antidiquark pair.

In the finite spatial volume of our lattice with extent $L \approx 2.90 \, \textrm{fm}$ we find two low-lying energy eigenstates, one around $60 \, \textrm{MeV}$ below the $D K$ threshold, the other slightly above the $D K$ threshold. The GEVP eigenvector components and the AMIAS coefficients and PDFs clearly indicate that the state below threshold, which corresponds to the $D_{s0}^\ast(2317)$ meson, is mostly of quark-antiquark type with only a small $D K$ component, while the state above threshold is a $D K$ scattering state. The tetraquark interpolating fields explored in this work turned out to be essentially irrelevant, when extracting the corresponding two energy levels, i.e.\ the couplings of the state below threshold to these interpolating fields is close to zero. We interpret this as indication that the $D_{s0}^\ast(2317)$ meson is mainly a quark-antiquark state and not a tetraquark, as discussed or proposed by various existing papers.

It is important to keep in mind that our computation was carried out for a single spatial volume and at quark masses different from those in the real world, in particular a $u$ and $d$ quark mass corresponding to a heavier pion, $m_\pi \approx 0.296 \, \textrm{GeV}$. We performed a scattering analysis using L\"uscher's method to determine the mass of the $D_{s0}^\ast(2317)$ in the infinite volume limit. We find this mass $51 \, \textrm{MeV}$ below the $D K$ threshold, rather close to our finite volume result as well as to the experimental value $45 \, \textrm{MeV}$. Thus we expect that our findings concerning the importance of various interpolating fields as well as the quark composition of the $D_{s0}^\ast(2317)$ meson will also apply to infinite volume and physical quark masses at least on a qualitative level. Of course, it would be worthwhile and interesting to perform similar computations at physical quark masses and for several volumes in the future, in particular to check the approximate independence of the GEVP eigenvector components or the AMIAS coefficients from the spatial volume.


\section{Acknowledgments}

J.F.\ acknowledges financial support by the PRACE Fifth and Sixth Implementation Phase (PRACE-5IP, PRACE-6IP) program of the European Commission under grant agreement No 730913 and No 823767. M.W.\ acknowledges funding by the Heisenberg Programme of the Deutsche Forschungsgemeinschaft (DFG, German Research Foundation) -- Projektnummer 399217702.

This work was supported in part by the Helmholtz International Center for FAIR within the framework of the LOEWE program launched by the State of Hesse.

Calculations on the LOEWE-CSC and on the on the FUCHS-CSC high-performance computer of the Frankfurt University were conducted for this research. We would like to thank HPC-Hessen, funded by the State Ministry of Higher Education, Research and the Arts, for programming advice.

Computations have been performed using the Chroma software library \cite{Edwards:2004sx} with a multigrid solver \cite{Babich:2010qb}.


\bibliography{Ds0}

\begin{thebibliography}{67}
\expandafter\ifx\csname natexlab\endcsname\relax\def\natexlab#1{#1}\fi
\expandafter\ifx\csname bibnamefont\endcsname\relax
  \def\bibnamefont#1{#1}\fi
\expandafter\ifx\csname bibfnamefont\endcsname\relax
  \def\bibfnamefont#1{#1}\fi
\expandafter\ifx\csname citenamefont\endcsname\relax
  \def\citenamefont#1{#1}\fi
\expandafter\ifx\csname url\endcsname\relax
  \def\url#1{\texttt{#1}}\fi
\expandafter\ifx\csname urlprefix\endcsname\relax\def\urlprefix{URL }\fi
\providecommand{\bibinfo}[2]{#2}
\providecommand{\eprint}[2][]{\url{#2}}

\bibitem[{\citenamefont{Aubert et~al.}(2003)}]{Aubert:2003fg}
\bibinfo{author}{\bibfnamefont{B.}~\bibnamefont{Aubert}} \bibnamefont{et~al.}
  (\bibinfo{collaboration}{BaBar}), \bibinfo{journal}{Phys. Rev. Lett.}
  \textbf{\bibinfo{volume}{90}}, \bibinfo{pages}{242001}
  (\bibinfo{year}{2003}), \eprint{hep-ex/0304021}.

\bibitem[{\citenamefont{Besson et~al.}(2003)}]{Besson:2003cp}
\bibinfo{author}{\bibfnamefont{D.}~\bibnamefont{Besson}} \bibnamefont{et~al.}
  (\bibinfo{collaboration}{CLEO}), \bibinfo{journal}{Phys. Rev.}
  \textbf{\bibinfo{volume}{D68}}, \bibinfo{pages}{032002}
  (\bibinfo{year}{2003}), \bibinfo{note}{[Erratum: Phys.\ Rev.\ D75, 119908
  (2007)]}, \eprint{hep-ex/0305100}.

\bibitem[{\citenamefont{Krokovny et~al.}(2003)}]{Krokovny:2003zq}
\bibinfo{author}{\bibfnamefont{P.}~\bibnamefont{Krokovny}} \bibnamefont{et~al.}
  (\bibinfo{collaboration}{Belle}), \bibinfo{journal}{Phys. Rev. Lett.}
  \textbf{\bibinfo{volume}{91}}, \bibinfo{pages}{262002}
  (\bibinfo{year}{2003}), \eprint{hep-ex/0308019}.

\bibitem[{\citenamefont{Tanabashi et~al.}(2018)}]{Tanabashi:2018oca}
\bibinfo{author}{\bibfnamefont{M.}~\bibnamefont{Tanabashi}}
  \bibnamefont{et~al.} (\bibinfo{collaboration}{Particle Data Group}),
  \bibinfo{journal}{Phys. Rev.} \textbf{\bibinfo{volume}{D98}},
  \bibinfo{pages}{030001} (\bibinfo{year}{2018}).

\bibitem[{\citenamefont{Godfrey and Isgur}(1985)}]{Godfrey:1985xj}
\bibinfo{author}{\bibfnamefont{S.}~\bibnamefont{Godfrey}} \bibnamefont{and}
  \bibinfo{author}{\bibfnamefont{N.}~\bibnamefont{Isgur}},
  \bibinfo{journal}{Phys. Rev.} \textbf{\bibinfo{volume}{D32}},
  \bibinfo{pages}{189} (\bibinfo{year}{1985}).

\bibitem[{\citenamefont{Godfrey and Kokoski}(1991)}]{Godfrey:1986wj}
\bibinfo{author}{\bibfnamefont{S.}~\bibnamefont{Godfrey}} \bibnamefont{and}
  \bibinfo{author}{\bibfnamefont{R.}~\bibnamefont{Kokoski}},
  \bibinfo{journal}{Phys. Rev.} \textbf{\bibinfo{volume}{D43}},
  \bibinfo{pages}{1679} (\bibinfo{year}{1991}).

\bibitem[{\citenamefont{Ebert et~al.}(2010)\citenamefont{Ebert, Faustov, and
  Galkin}}]{Ebert:2009ua}
\bibinfo{author}{\bibfnamefont{D.}~\bibnamefont{Ebert}},
  \bibinfo{author}{\bibfnamefont{R.~N.} \bibnamefont{Faustov}},
  \bibnamefont{and} \bibinfo{author}{\bibfnamefont{V.~O.}
  \bibnamefont{Galkin}}, \bibinfo{journal}{Eur. Phys. J.}
  \textbf{\bibinfo{volume}{C66}}, \bibinfo{pages}{197} (\bibinfo{year}{2010}),
  \eprint{0910.5612}.

\bibitem[{\citenamefont{Maiani et~al.}(2005)\citenamefont{Maiani, Piccinini,
  Polosa, and Riquer}}]{Maiani:2004vq}
\bibinfo{author}{\bibfnamefont{L.}~\bibnamefont{Maiani}},
  \bibinfo{author}{\bibfnamefont{F.}~\bibnamefont{Piccinini}},
  \bibinfo{author}{\bibfnamefont{A.~D.} \bibnamefont{Polosa}},
  \bibnamefont{and} \bibinfo{author}{\bibfnamefont{V.}~\bibnamefont{Riquer}},
  \bibinfo{journal}{Phys. Rev.} \textbf{\bibinfo{volume}{D71}},
  \bibinfo{pages}{014028} (\bibinfo{year}{2005}), \eprint{hep-ph/0412098}.

\bibitem[{\citenamefont{Bracco et~al.}(2005)\citenamefont{Bracco, Lozea,
  Matheus, Navarra, and Nielsen}}]{Bracco:2005kt}
\bibinfo{author}{\bibfnamefont{M.~E.} \bibnamefont{Bracco}},
  \bibinfo{author}{\bibfnamefont{A.}~\bibnamefont{Lozea}},
  \bibinfo{author}{\bibfnamefont{R.~D.} \bibnamefont{Matheus}},
  \bibinfo{author}{\bibfnamefont{F.~S.} \bibnamefont{Navarra}},
  \bibnamefont{and} \bibinfo{author}{\bibfnamefont{M.}~\bibnamefont{Nielsen}},
  \bibinfo{journal}{Phys. Lett.} \textbf{\bibinfo{volume}{B624}},
  \bibinfo{pages}{217} (\bibinfo{year}{2005}), \eprint{hep-ph/0503137}.

\bibitem[{\citenamefont{Dmitrasinovic}(2005)}]{Dmitrasinovic:2005gc}
\bibinfo{author}{\bibfnamefont{V.}~\bibnamefont{Dmitrasinovic}},
  \bibinfo{journal}{Phys. Rev. Lett.} \textbf{\bibinfo{volume}{94}},
  \bibinfo{pages}{162002} (\bibinfo{year}{2005}).

\bibitem[{\citenamefont{Ebert et~al.}(2011)\citenamefont{Ebert, Faustov, and
  Galkin}}]{Ebert:2010af}
\bibinfo{author}{\bibfnamefont{D.}~\bibnamefont{Ebert}},
  \bibinfo{author}{\bibfnamefont{R.~N.} \bibnamefont{Faustov}},
  \bibnamefont{and} \bibinfo{author}{\bibfnamefont{V.~O.}
  \bibnamefont{Galkin}}, \bibinfo{journal}{Phys. Lett.}
  \textbf{\bibinfo{volume}{B696}}, \bibinfo{pages}{241} (\bibinfo{year}{2011}),
  \eprint{1011.2677}.

\bibitem[{\citenamefont{Barnes et~al.}(2003)\citenamefont{Barnes, Close, and
  Lipkin}}]{Barnes:2003dj}
\bibinfo{author}{\bibfnamefont{T.}~\bibnamefont{Barnes}},
  \bibinfo{author}{\bibfnamefont{F.~E.} \bibnamefont{Close}}, \bibnamefont{and}
  \bibinfo{author}{\bibfnamefont{H.~J.} \bibnamefont{Lipkin}},
  \bibinfo{journal}{Phys. Rev.} \textbf{\bibinfo{volume}{D68}},
  \bibinfo{pages}{054006} (\bibinfo{year}{2003}), \eprint{hep-ph/0305025}.

\bibitem[{\citenamefont{Chen and Li}(2004)}]{Chen:2004dy}
\bibinfo{author}{\bibfnamefont{Y.-Q.} \bibnamefont{Chen}} \bibnamefont{and}
  \bibinfo{author}{\bibfnamefont{X.-Q.} \bibnamefont{Li}},
  \bibinfo{journal}{Phys. Rev. Lett.} \textbf{\bibinfo{volume}{93}},
  \bibinfo{pages}{232001} (\bibinfo{year}{2004}), \eprint{hep-ph/0407062}.

\bibitem[{\citenamefont{Du et~al.}(2018)\citenamefont{Du, Albaladejo,
  Fernández-Soler, Guo, Hanhart, Mei{\ss}ner, Nieves, and Yao}}]{Du:2017zvv}
\bibinfo{author}{\bibfnamefont{M.-L.} \bibnamefont{Du}},
  \bibinfo{author}{\bibfnamefont{M.}~\bibnamefont{Albaladejo}},
  \bibinfo{author}{\bibfnamefont{P.}~\bibnamefont{Fernández-Soler}},
  \bibinfo{author}{\bibfnamefont{F.-K.} \bibnamefont{Guo}},
  \bibinfo{author}{\bibfnamefont{C.}~\bibnamefont{Hanhart}},
  \bibinfo{author}{\bibfnamefont{U.-G.} \bibnamefont{Mei{\ss}ner}},
  \bibinfo{author}{\bibfnamefont{J.}~\bibnamefont{Nieves}}, \bibnamefont{and}
  \bibinfo{author}{\bibfnamefont{D.-L.} \bibnamefont{Yao}},
  \bibinfo{journal}{Phys. Rev.} \textbf{\bibinfo{volume}{D98}},
  \bibinfo{pages}{094018} (\bibinfo{year}{2018}), \eprint{1712.07957}.

\bibitem[{\citenamefont{Martinez~Torres
  et~al.}(2018)\citenamefont{Martinez~Torres, Oset, Prelovsek, and
  Ramos}}]{MartinezTorres:2017bdo}
\bibinfo{author}{\bibfnamefont{A.}~\bibnamefont{Martinez~Torres}},
  \bibinfo{author}{\bibfnamefont{E.}~\bibnamefont{Oset}},
  \bibinfo{author}{\bibfnamefont{S.}~\bibnamefont{Prelovsek}},
  \bibnamefont{and} \bibinfo{author}{\bibfnamefont{A.}~\bibnamefont{Ramos}},
  \bibinfo{journal}{PoS} \textbf{\bibinfo{volume}{Hadron2017}},
  \bibinfo{pages}{024} (\bibinfo{year}{2018}), \eprint{1712.09468}.

\bibitem[{\citenamefont{Albaladejo et~al.}(2018)\citenamefont{Albaladejo,
  Fernandez-Soler, Nieves, and Ortega}}]{Albaladejo:2018mhb}
\bibinfo{author}{\bibfnamefont{M.}~\bibnamefont{Albaladejo}},
  \bibinfo{author}{\bibfnamefont{P.}~\bibnamefont{Fernandez-Soler}},
  \bibinfo{author}{\bibfnamefont{J.}~\bibnamefont{Nieves}}, \bibnamefont{and}
  \bibinfo{author}{\bibfnamefont{P.~G.} \bibnamefont{Ortega}},
  \bibinfo{journal}{Eur. Phys. J.} \textbf{\bibinfo{volume}{C78}},
  \bibinfo{pages}{722} (\bibinfo{year}{2018}), \eprint{1805.07104}.

\bibitem[{\citenamefont{Guo et~al.}(2019)\citenamefont{Guo, Liu, Mei{\ss}ner,
  Oller, and Rusetsky}}]{Guo:2018tjx}
\bibinfo{author}{\bibfnamefont{Z.-H.} \bibnamefont{Guo}},
  \bibinfo{author}{\bibfnamefont{L.}~\bibnamefont{Liu}},
  \bibinfo{author}{\bibfnamefont{U.-G.} \bibnamefont{Mei{\ss}ner}},
  \bibinfo{author}{\bibfnamefont{J.~A.} \bibnamefont{Oller}}, \bibnamefont{and}
  \bibinfo{author}{\bibfnamefont{A.}~\bibnamefont{Rusetsky}},
  \bibinfo{journal}{Eur. Phys. J.} \textbf{\bibinfo{volume}{C79}},
  \bibinfo{pages}{13} (\bibinfo{year}{2019}), \eprint{1811.05585}.

\bibitem[{\citenamefont{van Beveren and Rupp}(2003)}]{vanBeveren:2003kd}
\bibinfo{author}{\bibfnamefont{E.}~\bibnamefont{van Beveren}} \bibnamefont{and}
  \bibinfo{author}{\bibfnamefont{G.}~\bibnamefont{Rupp}},
  \bibinfo{journal}{Phys. Rev. Lett.} \textbf{\bibinfo{volume}{91}},
  \bibinfo{pages}{012003} (\bibinfo{year}{2003}), \eprint{hep-ph/0305035}.

\bibitem[{\citenamefont{Kolomeitsev and Lutz}(2004)}]{Kolomeitsev:2003ac}
\bibinfo{author}{\bibfnamefont{E.~E.} \bibnamefont{Kolomeitsev}}
  \bibnamefont{and} \bibinfo{author}{\bibfnamefont{M.~F.~M.}
  \bibnamefont{Lutz}}, \bibinfo{journal}{Phys. Lett.}
  \textbf{\bibinfo{volume}{B582}}, \bibinfo{pages}{39} (\bibinfo{year}{2004}),
  \eprint{hep-ph/0307133}.

\bibitem[{\citenamefont{Hofmann and Lutz}(2004)}]{Hofmann:2003je}
\bibinfo{author}{\bibfnamefont{J.}~\bibnamefont{Hofmann}} \bibnamefont{and}
  \bibinfo{author}{\bibfnamefont{M.~F.~M.} \bibnamefont{Lutz}},
  \bibinfo{journal}{Nucl. Phys.} \textbf{\bibinfo{volume}{A733}},
  \bibinfo{pages}{142} (\bibinfo{year}{2004}), \eprint{hep-ph/0308263}.

\bibitem[{\citenamefont{Guo et~al.}(2018)\citenamefont{Guo, Heo, and
  Lutz}}]{Guo:2018kno}
\bibinfo{author}{\bibfnamefont{X.-Y.} \bibnamefont{Guo}},
  \bibinfo{author}{\bibfnamefont{Y.}~\bibnamefont{Heo}}, \bibnamefont{and}
  \bibinfo{author}{\bibfnamefont{M.~F.~M.} \bibnamefont{Lutz}},
  \bibinfo{journal}{Phys. Rev.} \textbf{\bibinfo{volume}{D98}},
  \bibinfo{pages}{014510} (\bibinfo{year}{2018}), \eprint{1801.10122}.

\bibitem[{\citenamefont{Zhu}(2008)}]{Zhu:2007wz}
\bibinfo{author}{\bibfnamefont{S.-L.} \bibnamefont{Zhu}},
  \bibinfo{journal}{Int. J. Mod. Phys.} \textbf{\bibinfo{volume}{E17}},
  \bibinfo{pages}{283} (\bibinfo{year}{2008}), \eprint{hep-ph/0703225}.

\bibitem[{\citenamefont{Chen et~al.}(2017)\citenamefont{Chen, Chen, Liu, Liu,
  and Zhu}}]{Chen:2016spr}
\bibinfo{author}{\bibfnamefont{H.-X.} \bibnamefont{Chen}},
  \bibinfo{author}{\bibfnamefont{W.}~\bibnamefont{Chen}},
  \bibinfo{author}{\bibfnamefont{X.}~\bibnamefont{Liu}},
  \bibinfo{author}{\bibfnamefont{Y.-R.} \bibnamefont{Liu}}, \bibnamefont{and}
  \bibinfo{author}{\bibfnamefont{S.-L.} \bibnamefont{Zhu}},
  \bibinfo{journal}{Rept. Prog. Phys.} \textbf{\bibinfo{volume}{80}},
  \bibinfo{pages}{076201} (\bibinfo{year}{2017}), \eprint{1609.08928}.

\bibitem[{\citenamefont{Boyle}(1997)}]{Boyle:1997aq}
\bibinfo{author}{\bibfnamefont{P.}~\bibnamefont{Boyle}}
  (\bibinfo{collaboration}{UKQCD}), \bibinfo{journal}{Nucl. Phys. Proc. Suppl.}
  \textbf{\bibinfo{volume}{53}}, \bibinfo{pages}{398} (\bibinfo{year}{1997}).

\bibitem[{\citenamefont{Boyle}(1998)}]{Boyle:1997rk}
\bibinfo{author}{\bibfnamefont{P.}~\bibnamefont{Boyle}}
  (\bibinfo{collaboration}{UKQCD}), \bibinfo{journal}{Nucl. Phys. Proc. Suppl.}
  \textbf{\bibinfo{volume}{63}}, \bibinfo{pages}{314} (\bibinfo{year}{1998}),
  \eprint{hep-lat/9710036}.

\bibitem[{\citenamefont{Lewis and Woloshyn}(2000)}]{Lewis:2000sv}
\bibinfo{author}{\bibfnamefont{R.}~\bibnamefont{Lewis}} \bibnamefont{and}
  \bibinfo{author}{\bibfnamefont{R.~M.} \bibnamefont{Woloshyn}},
  \bibinfo{journal}{Phys. Rev.} \textbf{\bibinfo{volume}{D62}},
  \bibinfo{pages}{114507} (\bibinfo{year}{2000}), \eprint{hep-lat/0003011}.

\bibitem[{\citenamefont{Hein et~al.}(2000)\citenamefont{Hein, Collins, Davies,
  Ali~Khan, Newton, Morningstar, Shigemitsu, and Sloan}}]{Hein:2000qu}
\bibinfo{author}{\bibfnamefont{J.}~\bibnamefont{Hein}},
  \bibinfo{author}{\bibfnamefont{S.}~\bibnamefont{Collins}},
  \bibinfo{author}{\bibfnamefont{C.~T.~H.} \bibnamefont{Davies}},
  \bibinfo{author}{\bibfnamefont{A.}~\bibnamefont{Ali~Khan}},
  \bibinfo{author}{\bibfnamefont{H.}~\bibnamefont{Newton}},
  \bibinfo{author}{\bibfnamefont{C.}~\bibnamefont{Morningstar}},
  \bibinfo{author}{\bibfnamefont{J.}~\bibnamefont{Shigemitsu}},
  \bibnamefont{and} \bibinfo{author}{\bibfnamefont{J.~H.} \bibnamefont{Sloan}},
  \bibinfo{journal}{Phys. Rev.} \textbf{\bibinfo{volume}{D62}},
  \bibinfo{pages}{074503} (\bibinfo{year}{2000}), \eprint{hep-ph/0003130}.

\bibitem[{\citenamefont{Bali}(2003)}]{Bali:2003jv}
\bibinfo{author}{\bibfnamefont{G.~S.} \bibnamefont{Bali}},
  \bibinfo{journal}{Phys. Rev.} \textbf{\bibinfo{volume}{D68}},
  \bibinfo{pages}{071501} (\bibinfo{year}{2003}), \eprint{hep-ph/0305209}.

\bibitem[{\citenamefont{Dougall et~al.}(2003)\citenamefont{Dougall, Kenway,
  Maynard, and McNeile}}]{Dougall:2003hv}
\bibinfo{author}{\bibfnamefont{A.}~\bibnamefont{Dougall}},
  \bibinfo{author}{\bibfnamefont{R.~D.} \bibnamefont{Kenway}},
  \bibinfo{author}{\bibfnamefont{C.~M.} \bibnamefont{Maynard}},
  \bibnamefont{and} \bibinfo{author}{\bibfnamefont{C.}~\bibnamefont{McNeile}}
  (\bibinfo{collaboration}{UKQCD}), \bibinfo{journal}{Phys. Lett.}
  \textbf{\bibinfo{volume}{B569}}, \bibinfo{pages}{41} (\bibinfo{year}{2003}),
  \eprint{hep-lat/0307001}.

\bibitem[{\citenamefont{di~Pierro et~al.}(2004)\citenamefont{di~Pierro,
  El-Khadra, Gottlieb, Kronfeld, Mackenzie, Menscher, Okamoto, and
  Simone}}]{diPierro:2003iw}
\bibinfo{author}{\bibfnamefont{M.}~\bibnamefont{di~Pierro}},
  \bibinfo{author}{\bibfnamefont{A.~X.} \bibnamefont{El-Khadra}},
  \bibinfo{author}{\bibfnamefont{S.~A.} \bibnamefont{Gottlieb}},
  \bibinfo{author}{\bibfnamefont{A.~S.} \bibnamefont{Kronfeld}},
  \bibinfo{author}{\bibfnamefont{P.~B.} \bibnamefont{Mackenzie}},
  \bibinfo{author}{\bibfnamefont{D.~P.} \bibnamefont{Menscher}},
  \bibinfo{author}{\bibfnamefont{M.}~\bibnamefont{Okamoto}}, \bibnamefont{and}
  \bibinfo{author}{\bibfnamefont{J.~N.} \bibnamefont{Simone}},
  \bibinfo{journal}{Nucl. Phys. Proc. Suppl.} \textbf{\bibinfo{volume}{129}},
  \bibinfo{pages}{328} (\bibinfo{year}{2004}), \eprint{hep-lat/0310045}.

\bibitem[{\citenamefont{Mohler and Woloshyn}(2011)}]{Mohler:2011ke}
\bibinfo{author}{\bibfnamefont{D.}~\bibnamefont{Mohler}} \bibnamefont{and}
  \bibinfo{author}{\bibfnamefont{R.~M.} \bibnamefont{Woloshyn}},
  \bibinfo{journal}{Phys. Rev.} \textbf{\bibinfo{volume}{D84}},
  \bibinfo{pages}{054505} (\bibinfo{year}{2011}), \eprint{1103.5506}.

\bibitem[{\citenamefont{Namekawa et~al.}(2011)}]{Namekawa:2011wt}
\bibinfo{author}{\bibfnamefont{Y.}~\bibnamefont{Namekawa}} \bibnamefont{et~al.}
  (\bibinfo{collaboration}{PACS-CS}), \bibinfo{journal}{Phys. Rev.}
  \textbf{\bibinfo{volume}{D84}}, \bibinfo{pages}{074505}
  (\bibinfo{year}{2011}), \eprint{1104.4600}.

\bibitem[{\citenamefont{Bali et~al.}(2011)}]{Bali:2011dc}
\bibinfo{author}{\bibfnamefont{G.}~\bibnamefont{Bali}} \bibnamefont{et~al.},
  \bibinfo{journal}{PoS} \textbf{\bibinfo{volume}{LATTICE2011}},
  \bibinfo{pages}{135} (\bibinfo{year}{2011}), \eprint{1108.6147}.

\bibitem[{\citenamefont{Bali et~al.}(2013)\citenamefont{Bali, Collins, and
  Perez-Rubio}}]{Bali:2012ua}
\bibinfo{author}{\bibfnamefont{G.}~\bibnamefont{Bali}},
  \bibinfo{author}{\bibfnamefont{S.}~\bibnamefont{Collins}}, \bibnamefont{and}
  \bibinfo{author}{\bibfnamefont{P.}~\bibnamefont{Perez-Rubio}},
  \bibinfo{journal}{J. Phys. Conf. Ser.} \textbf{\bibinfo{volume}{426}},
  \bibinfo{pages}{012017} (\bibinfo{year}{2013}), \eprint{1212.0565}.

\bibitem[{\citenamefont{Moir et~al.}(2013)\citenamefont{Moir, Peardon, Ryan,
  Thomas, and Liu}}]{Moir:2013ub}
\bibinfo{author}{\bibfnamefont{G.}~\bibnamefont{Moir}},
  \bibinfo{author}{\bibfnamefont{M.}~\bibnamefont{Peardon}},
  \bibinfo{author}{\bibfnamefont{S.~M.} \bibnamefont{Ryan}},
  \bibinfo{author}{\bibfnamefont{C.~E.} \bibnamefont{Thomas}},
  \bibnamefont{and} \bibinfo{author}{\bibfnamefont{L.}~\bibnamefont{Liu}},
  \bibinfo{journal}{JHEP} \textbf{\bibinfo{volume}{1305}}, \bibinfo{pages}{021}
  (\bibinfo{year}{2013}), \eprint{1301.7670}.

\bibitem[{\citenamefont{Kalinowski and Wagner}(2015)}]{Kalinowski:2015bwa}
\bibinfo{author}{\bibfnamefont{M.}~\bibnamefont{Kalinowski}} \bibnamefont{and}
  \bibinfo{author}{\bibfnamefont{M.}~\bibnamefont{Wagner}},
  \bibinfo{journal}{Phys. Rev.} \textbf{\bibinfo{volume}{D92}},
  \bibinfo{pages}{094508} (\bibinfo{year}{2015}), \eprint{1509.02396}.

\bibitem[{\citenamefont{Cichy et~al.}(2016)\citenamefont{Cichy, Kalinowski, and
  Wagner}}]{Cichy:2016bci}
\bibinfo{author}{\bibfnamefont{K.}~\bibnamefont{Cichy}},
  \bibinfo{author}{\bibfnamefont{M.}~\bibnamefont{Kalinowski}},
  \bibnamefont{and} \bibinfo{author}{\bibfnamefont{M.}~\bibnamefont{Wagner}},
  \bibinfo{journal}{Phys. Rev.} \textbf{\bibinfo{volume}{D94}},
  \bibinfo{pages}{094503} (\bibinfo{year}{2016}), \eprint{1603.06467}.

\bibitem[{\citenamefont{Cheung et~al.}(2016)\citenamefont{Cheung, O'Hara, Moir,
  Peardon, Ryan, Thomas, and Tims}}]{Cheung:2016bym}
\bibinfo{author}{\bibfnamefont{G.~K.~C.} \bibnamefont{Cheung}},
  \bibinfo{author}{\bibfnamefont{C.}~\bibnamefont{O'Hara}},
  \bibinfo{author}{\bibfnamefont{G.}~\bibnamefont{Moir}},
  \bibinfo{author}{\bibfnamefont{M.}~\bibnamefont{Peardon}},
  \bibinfo{author}{\bibfnamefont{S.~M.} \bibnamefont{Ryan}},
  \bibinfo{author}{\bibfnamefont{C.~E.} \bibnamefont{Thomas}},
  \bibnamefont{and} \bibinfo{author}{\bibfnamefont{D.}~\bibnamefont{Tims}}
  (\bibinfo{collaboration}{Hadron Spectrum}), \bibinfo{journal}{JHEP}
  \textbf{\bibinfo{volume}{12}}, \bibinfo{pages}{089} (\bibinfo{year}{2016}),
  \eprint{1610.01073}.

\bibitem[{\citenamefont{Chen and Chiu}(2017)}]{Chen:2017kxr}
\bibinfo{author}{\bibfnamefont{Y.-C.} \bibnamefont{Chen}} \bibnamefont{and}
  \bibinfo{author}{\bibfnamefont{T.-W.} \bibnamefont{Chiu}}
  (\bibinfo{collaboration}{TWQCD}), \bibinfo{journal}{Phys. Lett.}
  \textbf{\bibinfo{volume}{B767}}, \bibinfo{pages}{193} (\bibinfo{year}{2017}),
  \eprint{1701.02581}.

\bibitem[{\citenamefont{Mohler et~al.}(2013)\citenamefont{Mohler, Lang,
  Leskovec, Prelovsek, and Woloshyn}}]{Mohler:2013rwa}
\bibinfo{author}{\bibfnamefont{D.}~\bibnamefont{Mohler}},
  \bibinfo{author}{\bibfnamefont{C.~B.} \bibnamefont{Lang}},
  \bibinfo{author}{\bibfnamefont{L.}~\bibnamefont{Leskovec}},
  \bibinfo{author}{\bibfnamefont{S.}~\bibnamefont{Prelovsek}},
  \bibnamefont{and} \bibinfo{author}{\bibfnamefont{R.~M.}
  \bibnamefont{Woloshyn}}, \bibinfo{journal}{Phys. Rev. Lett.}
  \textbf{\bibinfo{volume}{111}}, \bibinfo{pages}{222001}
  (\bibinfo{year}{2013}), \eprint{1308.3175}.

\bibitem[{\citenamefont{Lang et~al.}(2014)\citenamefont{Lang, Leskovec, Mohler,
  Prelovsek, and Woloshyn}}]{Lang:2014yfa}
\bibinfo{author}{\bibfnamefont{C.~B.} \bibnamefont{Lang}},
  \bibinfo{author}{\bibfnamefont{L.}~\bibnamefont{Leskovec}},
  \bibinfo{author}{\bibfnamefont{D.}~\bibnamefont{Mohler}},
  \bibinfo{author}{\bibfnamefont{S.}~\bibnamefont{Prelovsek}},
  \bibnamefont{and} \bibinfo{author}{\bibfnamefont{R.~M.}
  \bibnamefont{Woloshyn}}, \bibinfo{journal}{Phys. Rev.}
  \textbf{\bibinfo{volume}{D90}}, \bibinfo{pages}{034510}
  (\bibinfo{year}{2014}), \eprint{1403.8103}.

\bibitem[{\citenamefont{Bali et~al.}(2017)\citenamefont{Bali, Collins, Cox, and
  Sch{\"a}fer}}]{Bali:2017pdv}
\bibinfo{author}{\bibfnamefont{G.~S.} \bibnamefont{Bali}},
  \bibinfo{author}{\bibfnamefont{S.}~\bibnamefont{Collins}},
  \bibinfo{author}{\bibfnamefont{A.}~\bibnamefont{Cox}}, \bibnamefont{and}
  \bibinfo{author}{\bibfnamefont{A.}~\bibnamefont{Sch{\"a}fer}}
  (\bibinfo{year}{2017}), \eprint{1706.01247}.

\bibitem[{\citenamefont{Liu et~al.}(2013)\citenamefont{Liu, Orginos, Guo,
  Hanhart, and Mei{\ss}ner}}]{Liu:2012zya}
\bibinfo{author}{\bibfnamefont{L.}~\bibnamefont{Liu}},
  \bibinfo{author}{\bibfnamefont{K.}~\bibnamefont{Orginos}},
  \bibinfo{author}{\bibfnamefont{F.-K.} \bibnamefont{Guo}},
  \bibinfo{author}{\bibfnamefont{C.}~\bibnamefont{Hanhart}}, \bibnamefont{and}
  \bibinfo{author}{\bibfnamefont{U.-G.} \bibnamefont{Mei{\ss}ner}},
  \bibinfo{journal}{Phys. Rev.} \textbf{\bibinfo{volume}{D87}},
  \bibinfo{pages}{014508} (\bibinfo{year}{2013}), \eprint{1208.4535}.

\bibitem[{\citenamefont{Moir et~al.}(2016)\citenamefont{Moir, Peardon, Ryan,
  Thomas, and Wilson}}]{Moir:2016srx}
\bibinfo{author}{\bibfnamefont{G.}~\bibnamefont{Moir}},
  \bibinfo{author}{\bibfnamefont{M.}~\bibnamefont{Peardon}},
  \bibinfo{author}{\bibfnamefont{S.~M.} \bibnamefont{Ryan}},
  \bibinfo{author}{\bibfnamefont{C.~E.} \bibnamefont{Thomas}},
  \bibnamefont{and} \bibinfo{author}{\bibfnamefont{D.~J.}
  \bibnamefont{Wilson}}, \bibinfo{journal}{JHEP} \textbf{\bibinfo{volume}{10}},
  \bibinfo{pages}{011} (\bibinfo{year}{2016}), \eprint{1607.07093}.

\bibitem[{\citenamefont{Darvish et~al.}(2019)\citenamefont{Darvish, Brett,
  Bulava, Fallica, Hanlon, H{\"o}rz, and Morningstar}}]{Darvish:2019oie}
\bibinfo{author}{\bibfnamefont{D.}~\bibnamefont{Darvish}},
  \bibinfo{author}{\bibfnamefont{R.}~\bibnamefont{Brett}},
  \bibinfo{author}{\bibfnamefont{J.}~\bibnamefont{Bulava}},
  \bibinfo{author}{\bibfnamefont{J.}~\bibnamefont{Fallica}},
  \bibinfo{author}{\bibfnamefont{A.}~\bibnamefont{Hanlon}},
  \bibinfo{author}{\bibfnamefont{B.}~\bibnamefont{H{\"o}rz}}, \bibnamefont{and}
  \bibinfo{author}{\bibfnamefont{C.}~\bibnamefont{Morningstar}}, in
  \emph{\bibinfo{booktitle}{{15th International Conference on Meson-Nucleon
  Physics and the Structure of the Nucleon (MENU 2019) Pittsburgh,
  Pennsylvania, USA, June 2-7, 2019}}} (\bibinfo{year}{2019}),
  \eprint{1909.07747}.

\bibitem[{\citenamefont{Alexandrou et~al.}(2015)\citenamefont{Alexandrou,
  Leontiou, Papanicolas, and Stiliaris}}]{Alexandrou:2014mka}
\bibinfo{author}{\bibfnamefont{C.}~\bibnamefont{Alexandrou}},
  \bibinfo{author}{\bibfnamefont{T.}~\bibnamefont{Leontiou}},
  \bibinfo{author}{\bibfnamefont{C.~N.} \bibnamefont{Papanicolas}},
  \bibnamefont{and}
  \bibinfo{author}{\bibfnamefont{E.}~\bibnamefont{Stiliaris}},
  \bibinfo{journal}{Phys. Rev.} \textbf{\bibinfo{volume}{D91}},
  \bibinfo{pages}{014506} (\bibinfo{year}{2015}), \eprint{1411.6765}.

\bibitem[{\citenamefont{Alexandrou et~al.}(2018)\citenamefont{Alexandrou,
  Berlin, Dalla~Brida, Finkenrath, Leontiou, and Wagner}}]{Alexandrou:2017itd}
\bibinfo{author}{\bibfnamefont{C.}~\bibnamefont{Alexandrou}},
  \bibinfo{author}{\bibfnamefont{J.}~\bibnamefont{Berlin}},
  \bibinfo{author}{\bibfnamefont{M.}~\bibnamefont{Dalla~Brida}},
  \bibinfo{author}{\bibfnamefont{J.}~\bibnamefont{Finkenrath}},
  \bibinfo{author}{\bibfnamefont{T.}~\bibnamefont{Leontiou}}, \bibnamefont{and}
  \bibinfo{author}{\bibfnamefont{M.}~\bibnamefont{Wagner}},
  \bibinfo{journal}{Phys. Rev.} \textbf{\bibinfo{volume}{D97}},
  \bibinfo{pages}{034506} (\bibinfo{year}{2018}), \eprint{1711.09815}.

\bibitem[{\citenamefont{Blossier et~al.}(2009)\citenamefont{Blossier,
  Della~Morte, von Hippel, Mendes, and Sommer}}]{Blossier:2009kd}
\bibinfo{author}{\bibfnamefont{B.}~\bibnamefont{Blossier}},
  \bibinfo{author}{\bibfnamefont{M.}~\bibnamefont{Della~Morte}},
  \bibinfo{author}{\bibfnamefont{G.}~\bibnamefont{von Hippel}},
  \bibinfo{author}{\bibfnamefont{T.}~\bibnamefont{Mendes}}, \bibnamefont{and}
  \bibinfo{author}{\bibfnamefont{R.}~\bibnamefont{Sommer}},
  \bibinfo{journal}{JHEP} \textbf{\bibinfo{volume}{0904}}, \bibinfo{pages}{094}
  (\bibinfo{year}{2009}), \eprint{0902.1265}.

\bibitem[{\citenamefont{Cheung et~al.}(2017)\citenamefont{Cheung, Thomas,
  Dudek, and Edwards}}]{Cheung:2017tnt}
\bibinfo{author}{\bibfnamefont{G.~K.~C.} \bibnamefont{Cheung}},
  \bibinfo{author}{\bibfnamefont{C.~E.} \bibnamefont{Thomas}},
  \bibinfo{author}{\bibfnamefont{J.~J.} \bibnamefont{Dudek}}, \bibnamefont{and}
  \bibinfo{author}{\bibfnamefont{R.~G.} \bibnamefont{Edwards}}
  (\bibinfo{collaboration}{Hadron Spectrum}), \bibinfo{journal}{JHEP}
  \textbf{\bibinfo{volume}{11}}, \bibinfo{pages}{033} (\bibinfo{year}{2017}),
  \eprint{1709.01417}.

\bibitem[{\citenamefont{Albanese et~al.}(1987)}]{Albanese:1987ds}
\bibinfo{author}{\bibfnamefont{M.}~\bibnamefont{Albanese}} \bibnamefont{et~al.}
  (\bibinfo{collaboration}{APE}), \bibinfo{journal}{Phys. Lett.}
  \textbf{\bibinfo{volume}{B192}}, \bibinfo{pages}{163} (\bibinfo{year}{1987}).

\bibitem[{\citenamefont{Gusken}(1990)}]{Gusken:1989qx}
\bibinfo{author}{\bibfnamefont{S.}~\bibnamefont{Gusken}},
  \bibinfo{journal}{Nucl. Phys. Proc. Suppl.} \textbf{\bibinfo{volume}{17}},
  \bibinfo{pages}{361} (\bibinfo{year}{1990}).

\bibitem[{\citenamefont{Jansen et~al.}(2008)\citenamefont{Jansen, Michael,
  Shindler, and Wagner}}]{Jansen:2008si}
\bibinfo{author}{\bibfnamefont{K.}~\bibnamefont{Jansen}},
  \bibinfo{author}{\bibfnamefont{C.}~\bibnamefont{Michael}},
  \bibinfo{author}{\bibfnamefont{A.}~\bibnamefont{Shindler}}, \bibnamefont{and}
  \bibinfo{author}{\bibfnamefont{M.}~\bibnamefont{Wagner}}
  (\bibinfo{collaboration}{ETM}), \bibinfo{journal}{JHEP}
  \textbf{\bibinfo{volume}{0812}}, \bibinfo{pages}{058} (\bibinfo{year}{2008}),
  \eprint{0810.1843}.

\bibitem[{\citenamefont{Aoki et~al.}(2009)}]{Aoki:2008sm}
\bibinfo{author}{\bibfnamefont{S.}~\bibnamefont{Aoki}} \bibnamefont{et~al.}
  (\bibinfo{collaboration}{PACS-CS}), \bibinfo{journal}{Phys. Rev.}
  \textbf{\bibinfo{volume}{D79}}, \bibinfo{pages}{034503}
  (\bibinfo{year}{2009}), \eprint{0807.1661}.

\bibitem[{\citenamefont{Abdel-Rehim et~al.}(2017)\citenamefont{Abdel-Rehim,
  Alexandrou, Berlin, Dalla~Brida, Finkenrath, and
  Wagner}}]{Abdel-Rehim:2017dok}
\bibinfo{author}{\bibfnamefont{A.}~\bibnamefont{Abdel-Rehim}},
  \bibinfo{author}{\bibfnamefont{C.}~\bibnamefont{Alexandrou}},
  \bibinfo{author}{\bibfnamefont{J.}~\bibnamefont{Berlin}},
  \bibinfo{author}{\bibfnamefont{M.}~\bibnamefont{Dalla~Brida}},
  \bibinfo{author}{\bibfnamefont{J.}~\bibnamefont{Finkenrath}},
  \bibnamefont{and} \bibinfo{author}{\bibfnamefont{M.}~\bibnamefont{Wagner}},
  \bibinfo{journal}{Comput. Phys. Commun.} \textbf{\bibinfo{volume}{220}},
  \bibinfo{pages}{97} (\bibinfo{year}{2017}), \eprint{1701.07228}.

\bibitem[{\citenamefont{Donnellan et~al.}(2011)\citenamefont{Donnellan,
  Knechtli, Leder, and Sommer}}]{Donnellan:2010mx}
\bibinfo{author}{\bibfnamefont{M.}~\bibnamefont{Donnellan}},
  \bibinfo{author}{\bibfnamefont{F.}~\bibnamefont{Knechtli}},
  \bibinfo{author}{\bibfnamefont{B.}~\bibnamefont{Leder}}, \bibnamefont{and}
  \bibinfo{author}{\bibfnamefont{R.}~\bibnamefont{Sommer}},
  \bibinfo{journal}{Nucl. Phys.} \textbf{\bibinfo{volume}{B849}},
  \bibinfo{pages}{45} (\bibinfo{year}{2011}), \eprint{1012.3037}.

\bibitem[{\citenamefont{Alexandrou et~al.}(2008)\citenamefont{Alexandrou,
  Papanicolas, and Stiliaris}}]{Alexandrou:2008bp}
\bibinfo{author}{\bibfnamefont{C.}~\bibnamefont{Alexandrou}},
  \bibinfo{author}{\bibfnamefont{C.~N.} \bibnamefont{Papanicolas}},
  \bibnamefont{and}
  \bibinfo{author}{\bibfnamefont{E.}~\bibnamefont{Stiliaris}},
  \bibinfo{journal}{PoS} \textbf{\bibinfo{volume}{LATTICE2008}},
  \bibinfo{pages}{099} (\bibinfo{year}{2008}), \eprint{0810.3982}.

\bibitem[{\citenamefont{Papanicolas and Stiliaris}(2012)}]{Papanicolas:2012sb}
\bibinfo{author}{\bibfnamefont{C.~N.} \bibnamefont{Papanicolas}}
  \bibnamefont{and} \bibinfo{author}{\bibfnamefont{E.}~\bibnamefont{Stiliaris}}
  (\bibinfo{year}{2012}), \eprint{arXiv:1205.6505}.

\bibitem[{\citenamefont{Michael}(1994)}]{Michael:1993yj}
\bibinfo{author}{\bibfnamefont{C.}~\bibnamefont{Michael}},
  \bibinfo{journal}{Phys. Rev.} \textbf{\bibinfo{volume}{D49}},
  \bibinfo{pages}{2616} (\bibinfo{year}{1994}), \eprint{hep-lat/9310026}.

\bibitem[{Note1()}]{Note1}
Note1, \bibinfo{note}{here and in the following we convert lattice results,
  which are obtained in units of the lattice spacing $a$, to $\protect \textrm
  {GeV}$ or $\protect \textrm {fm}$ by multiplying with appropriate powers of
  $a = 0.0907 \protect \tmspace +\thinmuskip {.1667em} \protect \textrm {fm}$.
  The error on the lattice spacing, $\Delta a = 0.0014 \protect \tmspace
  +\thinmuskip {.1667em} \protect \textrm {fm}$ is not taken into account.}

\bibitem[{\citenamefont{Ottnad et~al.}(2012)\citenamefont{Ottnad, Michael,
  Reker, Urbach, Michael, Reker, and Urbach}}]{Ottnad:2012fv}
\bibinfo{author}{\bibfnamefont{K.}~\bibnamefont{Ottnad}},
  \bibinfo{author}{\bibfnamefont{C.}~\bibnamefont{Michael}},
  \bibinfo{author}{\bibfnamefont{S.}~\bibnamefont{Reker}},
  \bibinfo{author}{\bibfnamefont{C.}~\bibnamefont{Urbach}},
  \bibinfo{author}{\bibfnamefont{C.}~\bibnamefont{Michael}},
  \bibinfo{author}{\bibfnamefont{S.}~\bibnamefont{Reker}}, \bibnamefont{and}
  \bibinfo{author}{\bibfnamefont{C.}~\bibnamefont{Urbach}}
  (\bibinfo{collaboration}{ETM}), \bibinfo{journal}{JHEP}
  \textbf{\bibinfo{volume}{11}}, \bibinfo{pages}{048} (\bibinfo{year}{2012}),
  \eprint{1206.6719}.

\bibitem[{\citenamefont{Michael et~al.}(2013)\citenamefont{Michael, Ottnad, and
  Urbach}}]{Michael:2013gka}
\bibinfo{author}{\bibfnamefont{C.}~\bibnamefont{Michael}},
  \bibinfo{author}{\bibfnamefont{K.}~\bibnamefont{Ottnad}}, \bibnamefont{and}
  \bibinfo{author}{\bibfnamefont{C.}~\bibnamefont{Urbach}}
  (\bibinfo{collaboration}{ETM}), \bibinfo{journal}{Phys. Rev. Lett.}
  \textbf{\bibinfo{volume}{111}}, \bibinfo{pages}{181602}
  (\bibinfo{year}{2013}), \eprint{1310.1207}.

\bibitem[{Note2()}]{Note2}
Note2, \bibinfo{note}{six terms in the truncated sum led to stable results for
  the energy differences $\protect \mathcal {E}_0$ to $\protect \mathcal
  {E}_5$, i.e.\ there is no significant change in the corresponding PDFs, when
  using more than six terms.}

\bibitem[{\citenamefont{L{\"u}scher}(1991)}]{Luscher:1990ux}
\bibinfo{author}{\bibfnamefont{M.}~\bibnamefont{L{\"u}scher}},
  \bibinfo{journal}{Nucl. Phys.} \textbf{\bibinfo{volume}{B354}},
  \bibinfo{pages}{531} (\bibinfo{year}{1991}).

\bibitem[{\citenamefont{Leskovec et~al.}(2019)\citenamefont{Leskovec, Meinel,
  Pflaumer, and Wagner}}]{Leskovec:2019ioa}
\bibinfo{author}{\bibfnamefont{L.}~\bibnamefont{Leskovec}},
  \bibinfo{author}{\bibfnamefont{S.}~\bibnamefont{Meinel}},
  \bibinfo{author}{\bibfnamefont{M.}~\bibnamefont{Pflaumer}}, \bibnamefont{and}
  \bibinfo{author}{\bibfnamefont{M.}~\bibnamefont{Wagner}},
  \bibinfo{journal}{Phys. Rev.} \textbf{\bibinfo{volume}{D100}},
  \bibinfo{pages}{014503} (\bibinfo{year}{2019}), \eprint{1904.04197}.

\bibitem[{\citenamefont{Briceno et~al.}(2018)\citenamefont{Briceno, Dudek, and
  Young}}]{Briceno:2017max}
\bibinfo{author}{\bibfnamefont{R.~A.} \bibnamefont{Briceno}},
  \bibinfo{author}{\bibfnamefont{J.~J.} \bibnamefont{Dudek}}, \bibnamefont{and}
  \bibinfo{author}{\bibfnamefont{R.~D.} \bibnamefont{Young}},
  \bibinfo{journal}{Rev. Mod. Phys.} \textbf{\bibinfo{volume}{90}},
  \bibinfo{pages}{025001} (\bibinfo{year}{2018}), \eprint{1706.06223}.

\bibitem[{\citenamefont{Edwards and Joo}(2005)}]{Edwards:2004sx}
\bibinfo{author}{\bibfnamefont{R.~G.} \bibnamefont{Edwards}} \bibnamefont{and}
  \bibinfo{author}{\bibfnamefont{B.}~\bibnamefont{Joo}}
  (\bibinfo{collaboration}{SciDAC, LHPC, UKQCD}), \bibinfo{journal}{Nucl. Phys.
  Proc. Suppl.} \textbf{\bibinfo{volume}{140}}, \bibinfo{pages}{832}
  (\bibinfo{year}{2005}), \eprint{hep-lat/0409003}.

\bibitem[{\citenamefont{Babich et~al.}(2010)\citenamefont{Babich, Brannick,
  Brower, Clark, Manteuffel, McCormick, Osborn, and Rebbi}}]{Babich:2010qb}
\bibinfo{author}{\bibfnamefont{R.}~\bibnamefont{Babich}},
  \bibinfo{author}{\bibfnamefont{J.}~\bibnamefont{Brannick}},
  \bibinfo{author}{\bibfnamefont{R.~C.} \bibnamefont{Brower}},
  \bibinfo{author}{\bibfnamefont{M.~A.} \bibnamefont{Clark}},
  \bibinfo{author}{\bibfnamefont{T.~A.} \bibnamefont{Manteuffel}},
  \bibinfo{author}{\bibfnamefont{S.~F.} \bibnamefont{McCormick}},
  \bibinfo{author}{\bibfnamefont{J.~C.} \bibnamefont{Osborn}},
  \bibnamefont{and} \bibinfo{author}{\bibfnamefont{C.}~\bibnamefont{Rebbi}},
  \bibinfo{journal}{Phys. Rev. Lett.} \textbf{\bibinfo{volume}{105}},
  \bibinfo{pages}{201602} (\bibinfo{year}{2010}), \eprint{1005.3043}.

\end{thebibliography}


\end{document}